%% file: isw_spie2007.tex
\documentclass[a4paper]{spie}  
\usepackage{graphicx}
\usepackage{subfigure}
\usepackage{amssymb}
\usepackage{hyperref}

\usepackage{ifpdf}

\include{thesis_macros}

\newcommand{\gpot}{\ensuremath{\Phi}}
\newcommand{\bias}{\ensuremath{b}}

\newcommand{\nd}{\ensuremath{\Delta^{\rm N}}}
\newcommand{\tp}{\ensuremath{\Delta^{\rm T}}}
\newcommand{\tpconj}{\ensuremath{\Delta^{{\rm T} \:\: \cconj}}}

\newcommand{\morphc}{\ensuremath{{D}}}
\newcommand{\morphe}{\ensuremath{{E}}}
\newcommand{\morphi}{\ensuremath{{I}}}
\newcommand{\morphs}{\ensuremath{{S}}}

\title{Detecting dark energy with wavelets on the sphere} 

\author{Jason D. McEwen
\skiplinehalf
Astrophysics Group, Cavendish Laboratory, Cambridge CB3 0HE, UK
}

\authorinfo{Further author information:\\ 
  E-mail: mcewen@mrao.cam.ac.uk, Telephone: +44 (0)1223 339 991}

\begin{document} 
  \maketitle 

\begin{abstract}
Dark energy dominates the energy density of our Universe, yet we know very little about its nature and origin.  Although strong evidence in support of dark energy is provided by the \cmbtext, the relic radiation of the Big Bang, in conjunction with either observations of supernovae or of the \lsstext\ of the Universe, the verification of dark energy by independent physical phenomena is of considerable interest.  We review works that, through a wavelet analysis on the sphere, independently verify the existence of dark energy by detecting the \iswtext\ effect.  The effectiveness of a wavelet analysis on the sphere is demonstrated by the highly statistically significant detections of dark energy that are made.  Moreover, the detection is used to constrain properties of dark energy.  A coherent picture of dark energy is obtained, adding further support to the now well established cosmological concordance model that describes our Universe.
\end{abstract}

\keywords{wavelets, spheres, cosmology, cosmic microwave background,
integrated Sachs-Wolfe effect}

\section{INTRODUCTION}
\label{sec:introduction} 

Dark energy represents the energy density of empty space, 
providing space with an intrinsic nature to expand.  The revolutionary idea that empty space contains energy was first proposed by Einstein, when he introduced the cosmological constant $\Lambda$ to reconcile his general relativistic model of the Universe with his prejudice towards a static universe.  Observational evidence\cite{hubble:1929} later contradicted Einstein's assumption, instead showing that the Universe is expanding.  Einstein subsequently referred to his introduction of the cosmological constant, which precluded him from predicting the expansion of the Universe, as his ``greatest blunder''.  Of course, only Einstein's greatest blunder could prove to be a remarkable insight which is today supported by strong observational evidence (although it was postulated for the wrong reasons initially).  We review in this paper a method that has been used to confirm independently the existence of dark energy and to constrain its properties.

The \iswtext\ (\isw) effect\cite{sachs:1967} provides an independent physical phenomenon that may be used to detect and probe dark energy.  This phenomenon predicts a secondary contribution to the anisotropies of the \cmbtext\ (\cmb), the relic radiation of the Big Bang.  \cmb\ photons undergo an energy shift due to the evolution of gravitational potentials as they travel towards us from emission shortly after the Big Bang.  This physical phenomenon is known as the \isw\ effect.  The secondary contribution to the \cmb\ due to the \isw\ effect is only present if the Universe deviates from matter domination due to either spatial curvature or dark energy.  Strong constraints have now been placed on the flatness of the Universe\cite{spergel:2006}, hence a detection of the \isw\ effect may be inferred as direct confirmation of the existence of dark energy.
It is difficult to isolate the \isw\ contribution to \cmb\ anisotropies directly.  Instead, as first proposed by Ref.~\citenum{crittenden:1996}, it is possible to detect the effect by searching for a correlation between the \cmb\ anisotropies and the \lsstext\ (\lss) of the Universe.  A positive large-scale correlation will be induced by the \isw\ effect as a consequence of decaying gravitational potentials due to dark energy. First attempts to detect the \isw\ effect using \cobedmrtext\ (\cobedmr) \cmb\ observations failed, concluding that greater sensitivity and resolution than that provided by \cobe\ were required\cite{boughn:2002}.  Fortunately, the \wmaptext\ (\wmap) mission soon provided suitable \cmb\ data.\cite{bennett:2003a,hinshaw:2006}  Correlations indicative of the \isw\ effect were first detected by Ref.~\citenum{boughn:2004} using the \wmap\ data and the \nvsstext\ (\nvss)\cite{condon:1998} radio galaxy survey, and have since been detected between various releases of the \wmap\ data and a large number of tracers of the \lss\ of the Universe.  

Many different analysis techniques have been employed to detect the \isw\ effect, each of which has its own merits and limitations.  We focus here on wavelet-based techniques.  Wavelets provide an ideal analysis tool to search for the \isw\ effect.  This is due to the localised nature of the physical phenomenon, in both scale and position on the sky, and the simultaneous scale and position localisation afforded by a wavelet analysis.  Since detections of the \isw\ effect are limited by the proportion of the sky observed, it is desirable to have as great a sky coverage as possible.  In this near full-sky setting the geometry of the celestial sphere on which observations are made should be taken into account.  Consequently, wavelet analyses on the sphere are required.  A variety of different methodologies have been proposed to construct a wavelet formalism on the sphere (see \eg\ Ref.~\citenum{antoine:1998} for a review).  Attempts to detect the \isw\ effect have predominately applied the methodlogy proposed by Ref.~\citenum{antoine:1998}.\footnote{A new wavelet construction on the sphere called needlets has also been used recently to detect the \isw\ effect\cite{pietrobon:2006}, although we do not focus on this work in any further detail here.}
The use of wavelets on the sphere to search for correlations indicative of the \isw\ effect was pioneered by Ref.~\citenum{vielva:2005}, using the axisymmetric spherical Mexican hat wavelet.  Correlated features induced by the \isw\ effect may not necessarily be rotationally invariant; indeed, it is known that statistically isotropic Gaussian random fields are characterised by local features that are not rotationally invariant\cite{barreiro:1997}.  The analysis performed by Ref.~\citenum{vielva:2005} was extended to directional wavelets by Ref.~\citenum{mcewen:2006:isw} in order to probe non-rotationally invariant features.  A fundamentally different analysis using wavelets on the sphere has been performed by Ref.~\citenum{mcewen:2007:isw2} recently.  This method employs steerable wavelets on the sphere to probe the morphology of local features.  All of these works examined the \wmap\ and \nvss\ data and made significant detections of the \isw\ effect, thereby confirming the existence of dark energy.  The effectiveness of the wavelet approach is highlighted in these analyses by the highly statistically significant detections of the \isw\ effect that have been made.  Moreover, in the first two of these works (Refs.~\citenum{vielva:2005,mcewen:2006:isw}) the detection of the \isw\ effect was used to constrain dark energy properties.  In this paper we review the aforementioned works that employ a wavelet analysis on the sphere to detect and constrain dark energy.

The remainder of this paper is organised as follows.  In \sectn{\ref{sec:cosmology}} we present the cosmological foundation on which the reviewed works are based.  The cosmological model that describes our Universe is discussed briefly, before the origin of the \isw\ effect is described.  In \sectn{\ref{sec:wavelets}} the wavelet methodology on the sphere applied to detect the \isw\ effect is presented.  Methods to detect the \isw\ are outlined in \sectn{\ref{sec:dark_energy}} and positive detections of the effect are presented.  Dark energy properties are also constrained in \sectn{\ref{sec:dark_energy}}.  Concluding remarks are made in \sectn{\ref{sec:conclusions}}.

\section{COSMOLOGY} 
\label{sec:cosmology}

Modern cosmology has converged on a concordance model only recently.  The so-called \lcdmtext\ (\lcdm) cosmological concordance model describes many independent observations of our Universe very well.  For example, the model describes the relic radiation of the Big Bang (\ie\ the \cmb), the \lss\ of our Universe and the accelerating expansion of our Universe apparent from observations of supernovae.  We describe this model very briefly here, before using it to explain the origin and observational implications of the \isw\ effect.

\subsection{Cosmological concordance model} 

The \lcdm\ model is characterised by a Universe consisting of ordinary baryonic matter, cold dark matter and dark energy.  Current estimates place the relative contributions of these components at 4\%, 22\% and 74\% of the energy density of the Universe respectively\cite{spergel:2006}.  Cold dark matter consists of non-relativistic, non-baryonic patricles that interact gravitationally only.  Although dark matter has yet to be observed directly, its presence has been inferred from a large range of observations, such as galaxy rotation curves\cite{rubin:1980}.  Dark energy represents the energy density of empty space and may be modelled by a cosmological fluid with negative pressure, acting as a repulsive force counteracting the attractive gravitational nature of matter.  In general, the dark energy fluid with pressure \pres\ and density \den\ has an equation of state $\pres=\w \den$, where $\w=-1$ corresponds to the case of a cosmological constant.  Evidence for dark energy is provided directly from observations of the statistical properties of the \cmb\ anisotropies, together with either observations of Type Ia Supernovae\cite{riess:1998,perlmutter:1999} or of the \lss\ of the Universe (\eg\ Ref.~\citenum{allen:2002}).  Although strong evidence in support of dark energy exists, a consistent model in the framework of particle physics is lacking.  Indeed, attempts to predict a cosmological constant using quantum field theory of the vacuum energy density arising from zero-point fluctuations predict a value that is too large by a factor of $\sim10^{120}$.  Dark energy dominates the energy density of our Universe and yet we know very little about its origin and nature.  The verification of dark energy by independent physical methods is therefore of considerable interest, hence the focus here on detecting dark energy through the \isw\ effect.  Moreover, independent methods may  provide more sensitive probes of certain properties of dark energy. 

Now that the fundamental components of our Universe have been described, we turn our attention to a dynamical view of the Universe.  Astronomical observations made by Hubble\cite{hubble:1929} (and since more accurately by others) showed that our Universe is expanding.  Moreover, distant regions of our Universe are receding from us at a rate proportional to their distance.  By reversing time and extrapolating into the past one recovers the Big Bang model, consisting of a universe that began from a singularity and expanded and evolved into the Universe observed today.  However, the Big Bang model as stated currently has a number of problems.  One of these problems is that the model contains no mechanism to generate the density perturbations required to explain the structure of the Universe that is observed today, such as galaxies and clusters of galaxies.  The theory of inflation provides a mechanism to solve this (and other) problem(s) of the original Big Bang model.  Inflation proposes that at a very early stage the Universe underwent a phase of rapid exponential expansion, increasing in size by many orders of magnitude.  The precise details of inflation are an active area of research, with many competing inflationary scenarios.  
In essence, a period of rapid expansion could occur if the Universe underwent a phase-transition, perhaps associated with symmetry breaking of the grand unified field, generating a false vacuum with a vacuum energy density acting like a cosmological repulsion.  The details of a scalar field with negative pressure which may drive such an expansion have yet to be identified.
The inflationary scenario provides a mechanism by which initial perturbations in the primordial fluid may be generated, thereby providing the seeds of structure formation.  
Quantum fluctuations in the very early Universe, or more precisely in the scalar field driving inflation, would be blown up to macroscopic scales by inflation, essentially freezing the quantum fluctuations into the Universe.  The resulting not quite perfectly homogeneous state may then evolve under gravitational instability to produce the structure of the Universe observed today. 

\begin{figure}[t]
\centering
\includegraphics[viewport= 0 0 2048 1050,clip=,width=80mm]
    {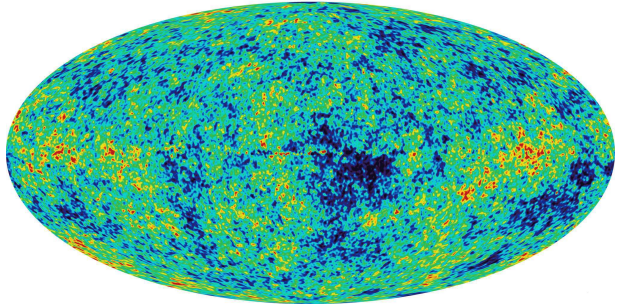}
\caption{Full-sky map of the \cmb\ anisotropies measured by \wmap\ (courtesy of the \wmap\ science team: \url{http://map.gsfc.nasa.gov/}).  The Mollweide projection is used here and subsequently to display data defined on the sphere.}
\label{fig:cmb}
\end{figure}

With the components and dynamics of the Universe now discussed, we are in a position to consider the final piece of the jigsaw required to understand the \isw\ effect:\ the \cmb.  
The temperature of the early Universe was sufficiently hot that photons had enough energy to ionise hydrogen.  Compton scattering tightly coupled photons to electrons, which were in turn coupled to baryons through electromagnetic interactions.  Scattering happened frequently and hence the mean free path of photons was extremely small.  
At this time the Universe therefore consisted of an opaque photon-baryon fluid.  As the Universe expanded, this primordial plasma cooled until the majority of photons no longer had sufficient energy to ionise hydrogen.  Consequently, the fraction of ionised electrons dropped, the photons decoupled from the baryons, and the Universe became essentially transparent to radiation.  This process, know as {\it recombination}, occurred when the temperature of the Universe fell to approximately 3000K, corresponding to time of approximately 400,000 years after the Big Bang (a small period relative to the age of the Universe, estimated at 13.7 billion years).  The photons were then free to propagate largely unhindered through space and may be observed today as the \cmb\ radiation.  The \cmb\ is highly isotropic, with anisotropies originating from primordial perturbations at a level of $10^{-5}$ only.  The \cmb\ was first observed by Penzias and Wilson in 1965\cite{penzias:1965} and since by many other ground and satellite based telescopes.  In \fig{\ref{fig:cmb}} we show the temperature anisotropies of the \cmb\ observered on the celestial sphere by \wmap.

\subsection{Integrated Sachs-Wolfe effect} 

As \cmb\ photons travel towards us from emission shortly after the Big Bang, they pass through gravitational potential wells due to the \lss\ of the Universe.  The photons suffer blue and red shifted as they fall into and out of potential wells respectively.  If the gravitational potential evolves during the photon propagation, then the
blue and red shifts do not cancel exactly and the photon undergoes an energy shift that produces a secondary contribution to the temperature anisotropy of the observed \cmb.  The \isw\ effect is the integrated sum of this energy shift along the photon path.  Any recent acceleration of the expansion of the Universe due to dark energy will cause local gravitational potentials to decay.  \cmb\ photons passing through over dense regions of decaying potential suffer blue shifts, resulting in a positive correlation between the induced anisotropy and the local matter distribution.  It can be shown\cite{sachs:1967} that the temperature fluctuation induced in the \cmb\ is given by
\begin{equation}
\label{eqn:isw}
\tp(\sa) =  \frac{\Delta T(\sa)}{T_0} = -2 \int \dx \ctime \: \dot{\gpot}(\ctime,\sa)
\spcend ,
\end{equation}
where $\Delta T$ is the induced temperature perturbation, $T_0$ is the mean temperature of the \cmb, \ctime\ is conformal time, \gpot\ is the gravitational potential and the dot represents a derivative with respect to conformal time.  
A point \sa\ on the sky is represented in spherical coordinates as $\sa=(\sas)$, with co-latitude \saa\ and longitude \sab.
The integral is computed over the photon path from emission to observation.
In a matter-dominated universe the gravitational potential remains constant with respect to conformal time (\ie\ $\dot{\gpot}=0$), thus there is no \isw\ effect (the expansion of the Universe exactly offsets the gravitational infall of over-dense regions).  However, if the Universe deviates from matter domination due to curvature or dark energy then an \isw\ effect is induced.  As mentioned previously, strong constraints have been placed on the flatness of the Universe, thus any \isw\ effect is a direct indication of dark energy.

It is difficult to isolate the \isw\ contribution to the \cmb\ anisotropies directly.  Instead, the effect may be detected by searching for a correlation between the \cmb\ and the \lss\ of the Universe\cite{crittenden:1996}.  The galaxy count distribution projected on to the sky $\delta^\ndlab(\z,\sa)$ may be used as a tracer of the local matter distribution $\delta(\z,\sa)$.  It is assumed that two relative fluctuations are related by the linear bias factor $\bias$, \ie\ \mbox{$\delta^\ndlab(\z,\sa) = \bias(\z) \, \delta(\z,\sa)$}, where \z\ is redshift and the bias is assumed to be redshift independent.  The galaxy source count fluctuation observed on the sky is therefore given by
\begin{equation}
\label{eqn:galaxy}
\nd (\sa) = \bias \: \int \dx z \: \frac{\dx N}{\dx \z} \: \delta(\z, \sa)
\spcend ,
\end{equation}
where $\dx N / \dx \z$ is the mean number of sources per steradian at redshift \z\ and the integral is performed from today to the epoch of recombination.

It is possible to derive the theoretical correlation between the \cmb\ temperature and galaxy count perturbations specified by \eqn{\ref{eqn:isw}} and \eqn{\ref{eqn:galaxy}} respectively.  The correlation is represented in harmonic space by the cross-power spectrum \clnttheo, defined by the ensemble average of the product of the spherical harmonic coefficients of the two signals observed on the sky:
\begin{equation}
\label{eqn:cltheo}
\opnexpv \nd_{\el\m} \: \tpconj_{\el\p \m\p} \clsexpv =
\kron{\el}{\el\p} \kron{\m}{\m\p} \:
\clnttheo
\spcend ,
\end{equation}
where 
$\Delta_{\el\m} = \opnexpv \shf{\el}{\m} | \Delta \clsexpv$ are the spherical harmonic coefficients of $\Delta(\sa)$, $\opnexpv \cdot | \cdot \clsexpv$ denotes the inner product on the sphere, $\shf{\el}{\m}$ are the spherical harmonic functions for multipole $\el\in\naturals$, $\m\in\integers$, $|\m|\leq\el$ and $\kron{i}{j}$ is the Kronecker delta symbol.  In writing the cross-correlation in this manner we implicitly assume that the galaxy density and \cmb\ random fields on the sphere are homogeneous and isotropic.  Representing the gravitational potential and the matter density perturbation in Fourier space and substituting \eqn{\ref{eqn:isw}} and \eqn{\ref{eqn:galaxy}} into \eqn{\ref{eqn:cltheo}}, it is possible to show (\eg\ Ref.~\citenum{nolta:2004}) that 
\begin{equation}
\clnttheo = 12 \pi \: \Omega_{\rm m} \: H_0{}^2
\int \frac{\dx k}{k^3} \:
\Delta_\delta^2(k) \:
F_\el^{\rm N}(k) \:
F_\el^{\rm T}(k)
\spcend ,
\end{equation}
where $\Omega_{\rm m}$ is the matter density, $H_0$ is the Hubble parameter, \mbox{$\Delta_\delta^2(k) = k^3 P_\delta(k)/2\pi^2$} is the logarithmic matter power spectrum,
$P_\delta(k)=\opnexpv |\delta(k)|^2 \clsexpv$ is the matter power spectrum and
the filter functions for the galaxy density and \cmb\ are given by
\begin{equation}
F_\el^{\ndlab}(k) = b \int \dx z \: \dndz \: D(z) \: \sbessel{\el}[k \eta(z)]
\end{equation}
and
\begin{equation}
F_\el^{\tplab}(k) = \int \dx z \: \dgdz \: \sbessel{\el}[k \eta(z)]
\end{equation}
respectively.  The integration required to compute $F_\el^\tplab(k)$ is performed over $z$ from zero to the epoch of recombination, whereas, in practice, the integration range for $F_\el^\ndlab(k)$ is defined by the source redshift distribution $\dx N / \dx \z$.  $D(\z)$ is the linear growth factor for the matter distribution: $\delta(\z,k)=D(\z) \delta(k)$, with $\delta(k) = \delta(0,k)$.  The function $g(\z) \equiv (1+\z)D(\z)$ is the linear growth suppression factor and $\sbessel{\el}(\cdot)$ is the spherical Bessel function.
We have represented in harmonic space the expected correlation between the galaxy source count and temperature fluctuations induced by the \isw\ effect.  We next turn our attention to wavelets on the sphere as a potential tool for detecting this predicted correlation.

\section{WAVELETS ON THE SPHERE} 
\label{sec:wavelets}

To perform a wavelet analysis of full-sky maps defined on the celestial sphere, Euclidean wavelet analysis must be extended to spherical geometry.  The \isw\ analyses that we review adopt the wavelet transform constructed by Ref.~\citenum{antoine:1998}.  This wavelet construction on the sphere is derived entirely from group theoretic principles, however the formalism has recently been reintroduced in an equivalent, practical and self-consistent approach that is independent of the original group theoretic framework.\cite{wiaux:2005}  We adopt this latter approach, with the extension to anisotropic dilations derived in Ref.~\citenum{mcewen:2006:fcswt}.\footnote{Although the ability to perform anisotropic dilations is of practical use, we do not achieve a wavelet basis in this setting since perfect synthesis of the original signal is not possible.  In any case, reconstruction is not required for this application.}  In order to perform the wavelet analysis on high resolution cosmological data sets it is essential to apply fast algorithms to compute the analysis.  The reviewed works make use of the fast algorithm derived by Ref.~\citenum{mcewen:2006:fcswt}, which is in turn based on the fast spherical convolution derived by Ref.~\citenum{wandelt:2001}.
We present here a brief overview of the wavelet transform on the sphere but refer the reader to Ref.~\citenum{wiaux:2006:review} for a more detailed review of this wavelet methodology and corresponding fast algorithms.

\subsection{Wavelet transform} 

The wavelet formalism on the sphere (\sphere) is constructed in an analogous manner to the construction of Euclidean wavelets.  Affine transformations, such as translations and dilations, must be defined on the sphere.  A wavelet basis on the sphere may then be constructed from affine transformations of a mother wavelet defined on the sphere.  The wavelet transform is given by the projection of a function on to this basis, where the mother wavelet must satisfy an admissibility relation to ensure that perfect synthesis of the original function from its wavelet coefficients is possible.  In the most general setting, the projection is given by a directional spherical convolution.  For a given scale, the wavelet coefficients therefore live in the rotation group \sothree.

\begin{figure}
\centerline{
  \includegraphics[width=86mm]{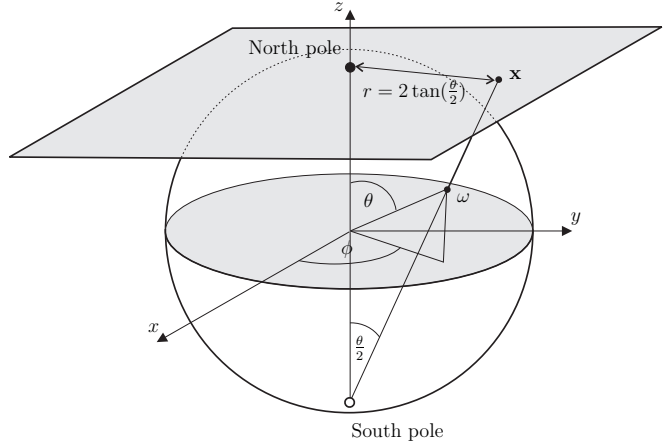}}
\caption{Stereographic projection of the sphere on to the plane.}
\label{fig:stereographic_projection}
\end{figure}

In order to construct the wavelet framework on the sphere following the methodology outline above, it is necessary to extend the affine transformations of translation and dilation to the sphere.  The natural extension of translations to the sphere are rotations.  These are characterised by the elements of the rotation group \sothree, which we parameterise in terms of the three Euler angles $\eul=(\euls)$.\footnote{We adopt the $zyz$ Euler convention corresponding to the rotation of a physical body in a \emph{fixed} co-ordinate system about the $z$, $y$ and $z$ axes by \eulc, \eulb\ and \eula\ respectively.}  The rotation of a function $\sky$ on the sphere is defined by
\begin{equation}
[\rot(\rho) \sky](\sa) = \sky(\rho^{-1} \sa), \quad \rho \in \sothree 
\spcend .
\end{equation}
The extension of the dilation operator to the sphere is not so simple and various operators have been proposed.  Refs.~\citenum{antoine:1998,wiaux:2005} define a dilation operator that may be conceptualised through an association with the tangent plane at the north pole of the sphere, using the stereographic projection to map a function defined on the sphere to the tangent plane.  The stereographic projection is defined by projecting a point on the sphere to a point on the tangent plane at the north pole, by casting a ray through the point and the south pole.  The point on the sphere is mapped on to the intersection of this ray and the tangent plane (see \fig{\ref{fig:stereographic_projection}}).  We denote the stereographic projection operator by $\spo$, with inverse $\spo^{-1}$, where these operators include the appropriate normalisation term to ensure that the \ltwo-norm of functions on the sphere and tangent plane is preserved.
In this formalism dilations on the sphere are constructed by first
lifting the sphere to the plane by the stereographic projection,
followed by the usual Euclidean dilation in the plane, before
re-projecting back on to the sphere.  We adopt the generalisation to anisotropic dilations on the sphere\cite{mcewen:2006:fcswt}.
The spherical dilation operator $\dil(\scaleab)$ is defined as the conjugation by \spo\ of
the anisotropic Euclidean dilation $d(\scaleab)$ on the tangent plane at the north pole:
\begin{equation}
\dil(\scaleab) \equiv \spo^{-1} \, \dilsmall(\scaleab) \, \spo
\spcend ,
\end{equation}
where the Euclidean dilation operator $d(\scaleab)$ has the appropriate normalisation factor to ensure that the \ltwo-norm is preserved.  Consequently, the \ltwo-norm of functions is preserved by the
spherical dilation as both the stereographic projection and
Euclidean dilation operators preserve the norm of functions.
The dilation of a function on the sphere $f\in\ltwo(\sphere)$ is given by
\begin{equation}
[\dil(\scaleab) f](\sa)
= [\cocycle(\scaleab,\sas)]^{1/2} \: f(\sa_{1/\scalea,1/\scaleb})
\spcend ,
\end{equation}
where $\sa_\scaleab=(\saa_\scaleab,\sab_\scaleab)$,
$
\tan(\saa_\scaleab/2) = 
\tan(\saa/2)
\sqrt{\scalea^2 \cos^2{\sab} + \scaleb^2 \sin^2{\sab}}
$
and 
$\tan(\sab_\scaleab) = 
\frac{\scaleb}{\scalea}
\tan(\sab)
$.
For the case where $\scalea=\scaleb$ the anisotropic
dilation reduces to the usual isotropic case defined by 
$\tan(\saa_\scalea/2) = \scalea \tan(\saa/2)$ and
$\sab_\scalea=\sab$.  
The $\cocycle(\scaleab,\sas)$ cocycle term follows from the factors
introduced in the stereographic projection of functions to preserve
the $\ltwo$-norm.  The cocycle of an
anisotropic spherical dilation is defined by
\begin{equation}
\cocycle(\scaleab, \sas) \equiv
\frac{4 \scalea^3 \scaleb^3}
{ 
\left(
A_{-}\cos\saa + A_{+}
\right)^2
}
\spcend ,
\end{equation}
where 
\begin{displaymath}
A_\pm = \scalea^2\scaleb^2 \pm \scalea^2 \sin^2\sab \pm \scaleb^2 \cos^2\sab
\spcend .
\end{displaymath}
For the case where $\scalea=\scaleb$ the anisotropic cocycle reduces
to the usual isotropic cocycle.
Although the ability to perform anisotropic dilations is of practical use, one does not achieve a wavelet basis in this setting. 
In the anisotropic setting a bounded admissibility integral cannot be determined (even in the plane), thus the synthesis of a signal from its coefficients cannot be performed.  Nevertheless, this is not of concern for the \isw\ analyses discussed herein since synthesis is not required.
The projection of a signal on to basis functions undergoing anisotropic dilations may be performed in an analogous manner to the following discussion of the wavelet transform.  However, since these basis functions are not wavelets we restrict the following discussion to isotropic dilations.

A wavelet basis on the sphere may now be constructed from rotations and
isotropic dilations (where \mbox{$\scalea=\scaleb$}) of a mother spherical wavelet $\wavi \in \ltwo(\sphere)$. 
The corresponding wavelet family 
$\{ \wavi_{\scalea,\rho} \equiv \rot(\eul) \dil(\scalea,\scalea) \wavi :
 {\rho \in \sothree,} $ \linebreak ${\scalea \in \reals_{\ast}^{+}} \}$
provides an over-complete set of functions in $\ltwo(\sphere)$. 
The wavelet transform on the sphere of a square integrable function $\sky \in \ltwo(\sphere)$ is given
by the projection on to each wavelet basis function in the usual manner:
\begin{equation}
\skywavi(\scalea, \eul) 
\equiv
\int_{\sphere}
\dmu{\sa} \:
\sky(\sa) \:
\wavi_{\scalea,\eul}^\cconj(\sa)
\spcend ,
\label{eqn:cswti}
\end{equation}
where $\dmu{\sa}=\sin\saa \dx \saa \dx \sab$ is the usual invariant measure on the sphere.
The transform is general in the sense that all orientations in the
rotation group \sothree\ are considered, thus directional structure is
naturally incorporated.  It is important to note, however, that only
\emph{local} directions make any sense on \sphere.  There is no global
way of defining directions on the sphere\footnote{There is no
differentiable vector field of constant norm on the sphere and hence
no global way of defining directions.} -- there will always be some
singular point where the definition fails.  
Although it is not of relevance to the \isw\ analyses reviewed herein, for completeness we also state the synthesis of a signal on the sphere from its wavelet coefficients:
  \begin{equation}
  \sky(\sa) = 
  \int_{\sothree} \deul{\eul}
  \int_0^\infty \frac{\dx\scalea}{\scalea^3} \:\:
  \skywavi(\scalea, \eul) \:
  [\rot(\eul) \Lopi \wavi_\scalea](\sa)
  \spcend ,
  \end{equation}
where $\deul{\eul}=\sin\eulb \dx \eula \dx \eulb \dx \eulc$ is the invariant measure on the rotation group \sothree\ and
the \Lopi\ operator is defined by the action
$
\shc{(\Lopi g)}{\el}{\m} \equiv \shc{g}{\el}{\m} / \admissCli 
$
on the spherical harmonic coefficients of functions $g \in \ltwo(\sphere)$.
In order to ensure the perfect reconstruction of a signal synthesised from its wavelet coefficients, the admissibility condition 
  \begin{equation}
  \label{eqn:admiss_full}
  0 <
  \admissCli \equiv
  \frac{8\pi^2}{2\el+1}
  \sum_{\m=-\el}^\el 
  \int_0^\infty
  \frac{\dx\scalea}{\scalea^3}
  \mid \shc{(\wavi_\scalea)}{\el}{\m} \mid^2
  < \infty
  \end{equation}
must hold for all $\el \in \naturals$, where $\shc{(\wavi_\scalea)}{\el}{\m}$ are the spherical harmonic coefficients of $\wavi_\scalea$.\cite{wiaux:2005}.

\subsection{Correspondence principle} 

The correspondence principle between spherical and Euclidean wavelets
states that the inverse stereographic projection of an
\emph{admissible} wavelet on the plane yields an 
\emph{admissible} wavelet on the sphere.\cite{wiaux:2005}.
Mother spherical wavelets may therefore be constructed from the projection
of mother Euclidean wavelets defined on the plane: 
\begin{equation}
\label{eqn:wav_proj}
\wavi(\sa) = [\spo^{-1}\wavi_{\reals^2}](\sa)
\spcend ,
\end{equation}
where $\wavi_{\reals^2} \in \ltwo(\reals^2)$ is an admissible wavelet in the plane.
Directional spherical wavelets may be naturally constructed in this 
setting -- they are simply the projection of directional Euclidean planar
wavelets on to the sphere.
In \fig{\ref{fig:wavelets}} we illustrate the three spherical wavelets that have been used to detect the \isw\ effect: the spherical Mexican
hat wavelet (\smhw); the spherical butterfly wavelet\footnote{Sometimes referred to as the spherical first Gaussian derivative wavelet.} (\sbw); and
the spherical second Gaussian derivative wavelet (\stwogdw).
Each spherical wavelet is constructed by the stereographic projection
of the corresponding Euclidean wavelet on to the sphere.  The \smhw\ is proportional to the Laplacian of a Gaussian,
the \sbw\ is proportional to the first partial derivative of a Gaussian in one direction and the \stwogdw\ is proportional to the second partial derivative of a Gaussian in one direction.  See Ref.~\citenum{wiaux:2006:review} for formal definitions of these wavelets.

\newlength{\wavplotwidth}
\setlength{\wavplotwidth}{40mm}

\begin{figure}
\centering
\mbox{
  \subfigure[\smhw]{\includegraphics[viewport= 0 -10 350 400,width=\wavplotwidth,clip]{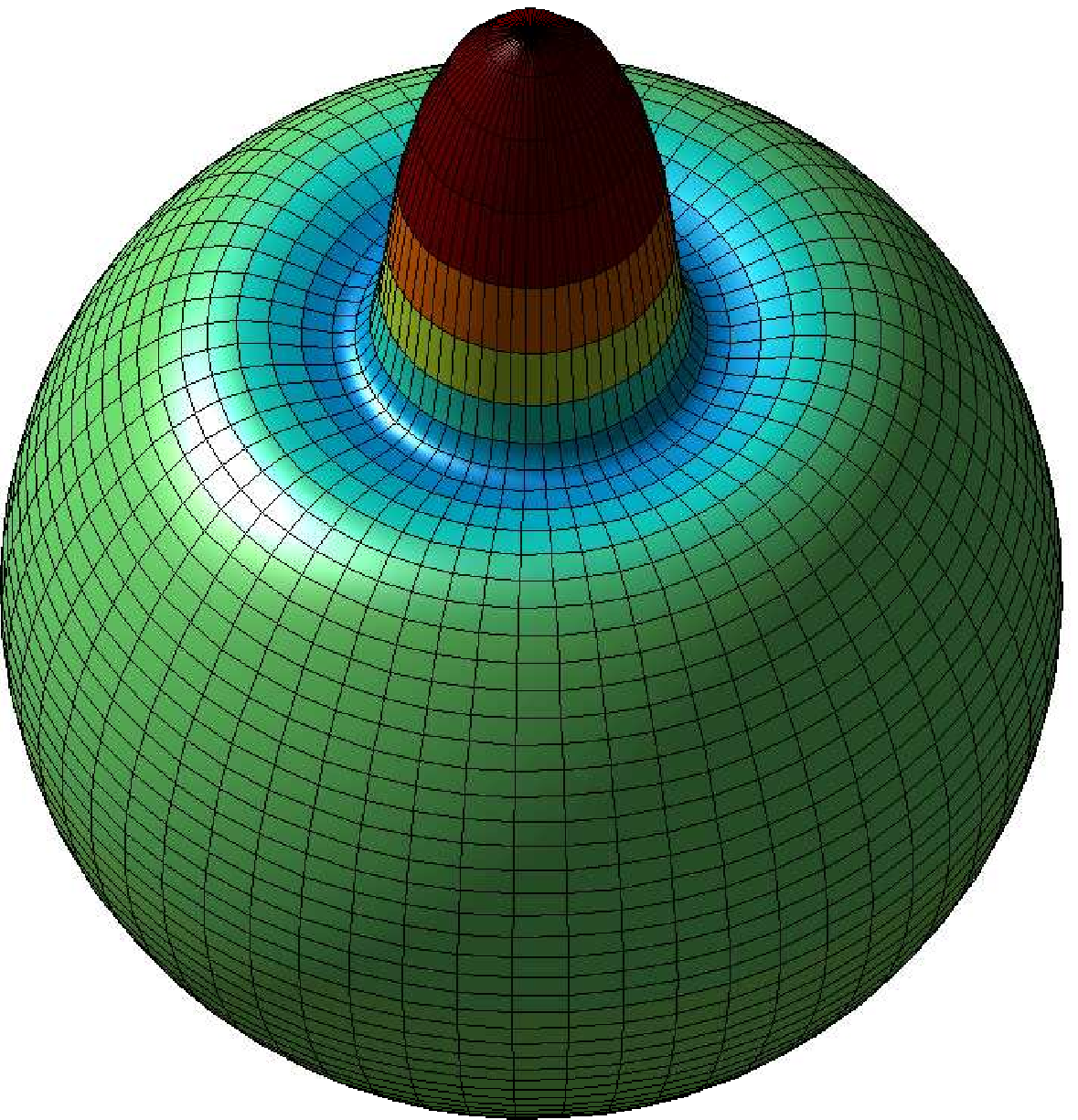}} \quad \quad
  \subfigure[\sbw]{\includegraphics[viewport= 0 -10 350 400,width=\wavplotwidth,clip]{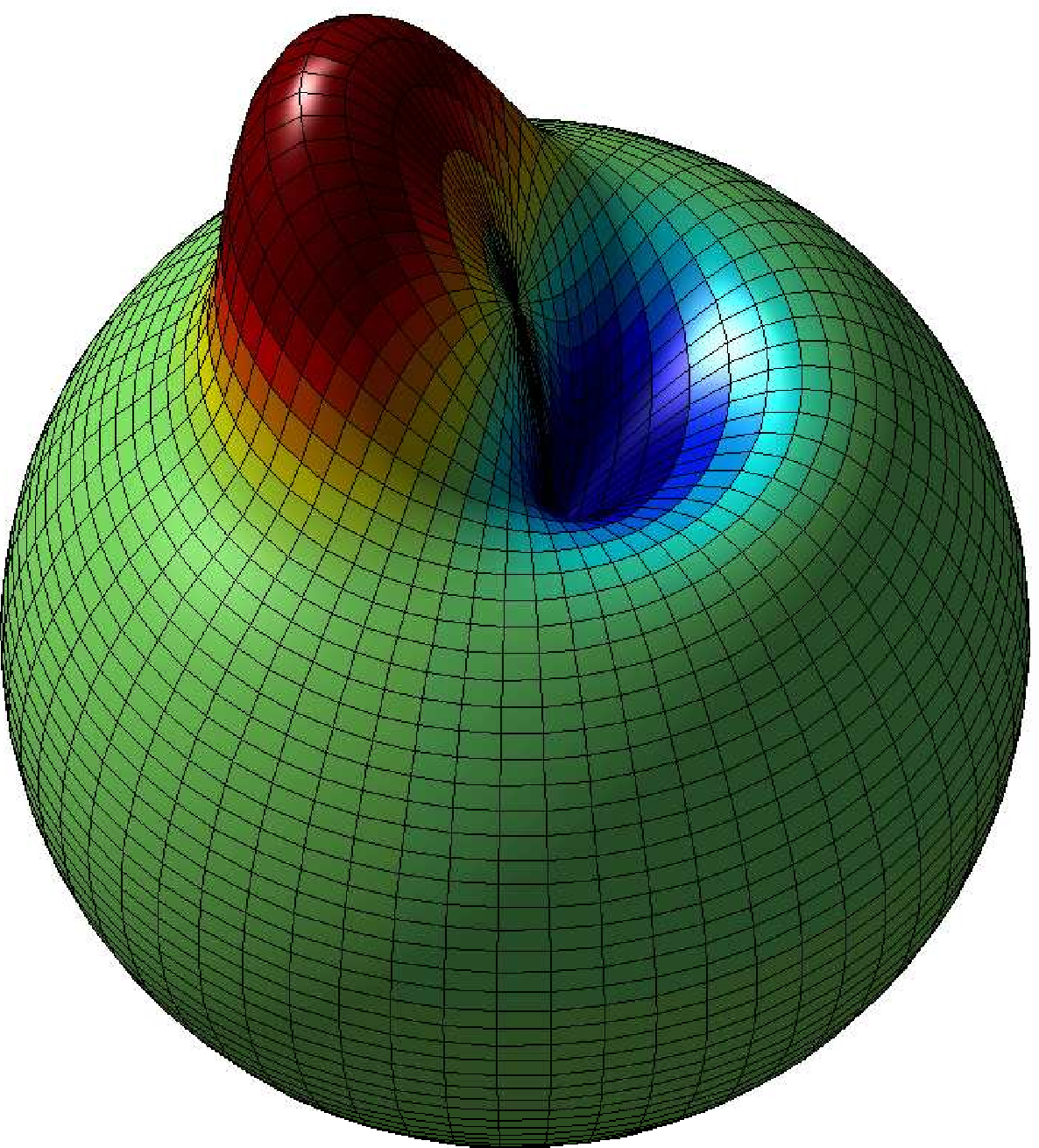}} \quad \quad
  \subfigure[\stwogdw]{\includegraphics[viewport= 0 -10 335 400,width=\wavplotwidth,clip]{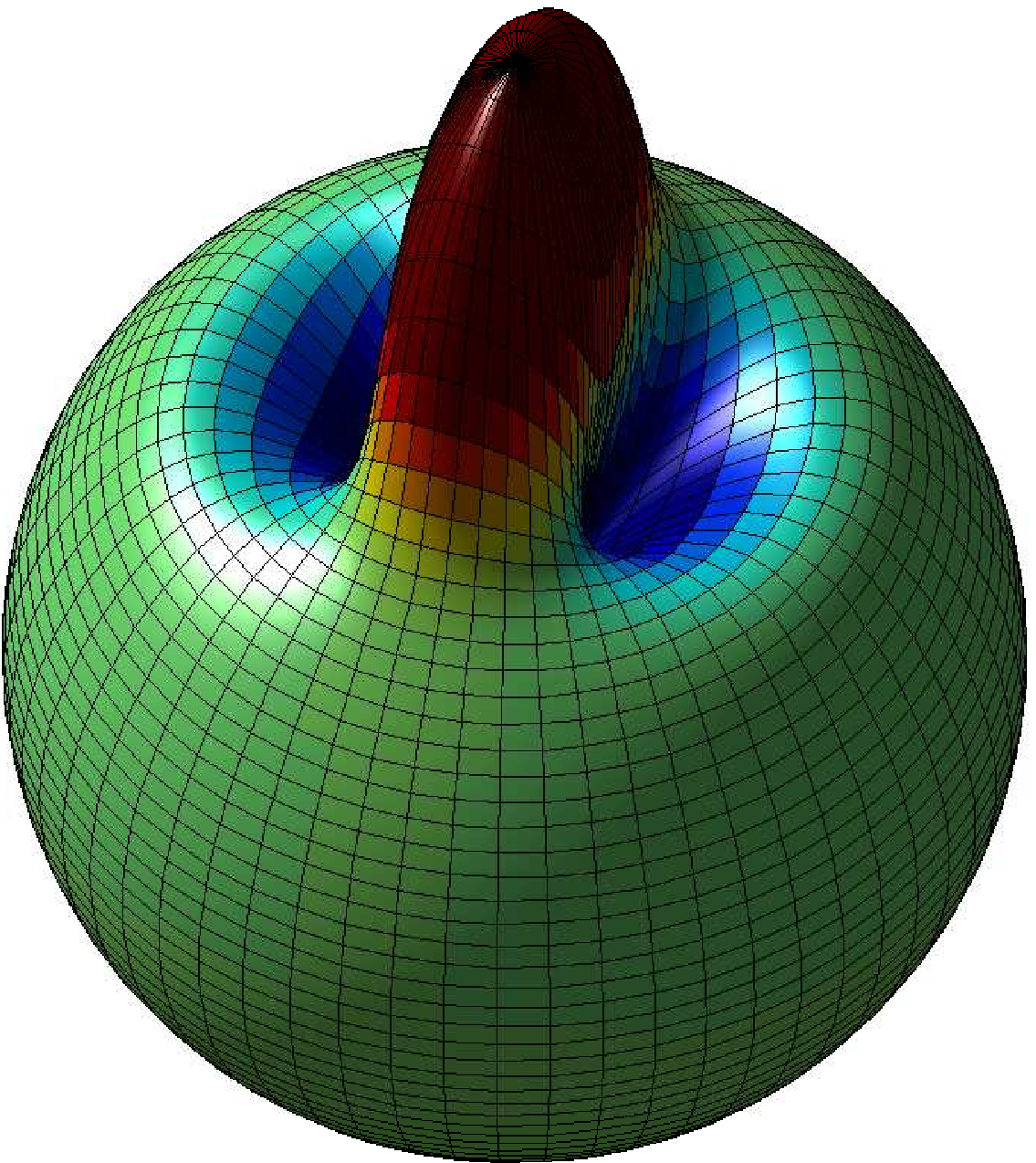}}
}
\caption{Spherical wavelets at scale $\scalea=\scaleb=0.2$.}
\label{fig:wavelets}
\end{figure}

\subsection{Steerability} 

The wavelet transform outlined above is inherently directional (since it lives in the rotation group \sothree), allowing for the construction of both axisymmetric and directional wavelets (an \emph{axisymmetric} wavelet is invariant under rotations around itself; all wavelets that are not axisymmetric, we refer to as \emph{directional}).  For example, the \smhw\ (\fig{\ref{fig:wavelets}a}) is axisymmetric, whereas the \sbw\ (\fig{\ref{fig:wavelets}b}) and the \stwogdw\ (\fig{\ref{fig:wavelets}c}) are directional.  Certain classes of directional wavelets are said to be \emph{steerable}.  A steerable wavelet for any arbitrary rotation about itself by \eulc\ may be represented by a weighted linear combination of non-rotated basis wavelets.  In some cases the basis wavelets are simply rotated versions of the steerable wavelet itself.  Due to the linearity of the wavelet transform this property may be transferred to the wavelet coefficients.  For steerable wavelets it is possible to compute the wavelet coefficients for any continuous orientation \eulc\ from a weighted sum of wavelet coefficients computed using the non-rotated basis wavelets:
\begin{equation}
{{W}}_\wavi^\sky(\scalea, \euls)
= \sum_{m=1}^{M} k_m(\eulc) {{W}}_{\wavi_m}^\sky(\scalea, \eula, \eulb, 0)
\spcend,
\end{equation}
where ${\wavi_m}$ are the $M$ basis wavelets and $k_m(\eulc)$ are the interpolating functions.  Both the \sbw\ and the \stwogdw\ are steerable, requiring $M=2$ and $M=3$ basis wavelets respectively.  For a more in-depth discussion of steerable wavelets on the sphere, with specific examples and interpolating functions specified, see Refs.~\citenum{wiaux:2005,wiaux:2006:review}.

The ability of steerable wavelets to probe all continuous orientations mean that it is possible to extract the single orientation corresponding to the most dominant feature at each position on the sphere.  It addition to the orientation of this feature, it is also possible to compute other morphological measures of the feature.  For example, the \stwogdw\ may also be used to compute measures of the signed-intensity and elongation of local features.\cite{mcewen:2007:isw2}  We denote the morphological measures computed at each position on the sphere as \morphi, \morphc\ and \morphe\ for the signed-intensity, orientation and elongation respectively.  By allowing one to probe the morphology of local features, steerability facilitates an additional approach that may be used to probe dark energy.

\section{DETECTING AND CONSTRAINING DARK ENERGY} 
\label{sec:dark_energy}

If dark energy exists, a correlation between the \cmb\ and the \lss\ of the Universe is predicted by the \isw\ effect.  In the absence of dark energy no correlation should exist.  To verify the existence of dark energy one simply needs to detect a correlation between the \cmb\ and a tracer of the \lss, such as the nearby radio galaxy distribution (and rule out other sources for any observed correlation).  Wavelets are used to search for this correlation due to the ability to simultaneously localise signal components in scale and position, thereby improving the sensitivity of the analysis.  In the remainder of this section we discuss the analyses performed in Refs.~\citenum{vielva:2005,mcewen:2006:isw,mcewen:2007:isw2} and present the results obtained.  Firstly, we describe the methods used, detailing the wavelet estimators applied, the data examined and the analysis procedures followed.  We then present the detections of dark energy that are made and the constraints that are placed on the properties of dark energy.

\subsection{Wavelet estimators} 

The correlation between the \cmb\ and \lss\ data is computed in wavelet space by
\begin{equation}
\label{eqn:covest}
X_{\morphs_i}^{\ndlab \tplab}(\scaleavect) = \frac{1}{N_{\rm p}} 
\sum_{\eul} 
\morphs_i^\ndlab(\scaleavect,\eul) \: \morphs_i^\tplab(\scaleavect, \eul)
-
\bar{\morphs}_i^\ndlab(\scaleavect) \: \bar{\morphs}_i^\tplab(\scaleavect) 
\spcend ,
\end{equation}
where $\morphs_i = \{
W,\:
\morphi,\:
\morphc,\:
\morphe 
\}$
is the term correlated (wavelet coefficient or morphological signed-intensity, orientation or elongation respectively), with superscript \ndlab\ or \tplab\ representing respectively the \lss\ and \cmb\ data and the bar denotes a mean computed over the sky (note that the correlated terms are computed at $N_{\rm p}$ discrete points only).  
The scale parameter is given by $\scaleavect=(\scaleab)^{\rm T}$, which is represented by $\scalea$ only when an isotropic analysis is performed.  In the absence of an \isw\ effect the \cmb\ and \lss\ data should be independent and none of the correlation estimators defined by \eqn{\ref{eqn:covest}} should exhibit a significant deviation from zero.
It is possible to define estimators of the correlation in real and harmonic space also.  However, it has been shown that the wavelet estimator is optimal for a large range of scales\cite{vielva:2005}.  The predicted effectiveness of various wavelets for detecting the \isw\ effect has also been compared, concluding that the wavelets that have been adopted are expected to perform well.\cite{mcewen:2006:isw}

\subsection{Data and simulations} 

The three-year release of the \wmap\ data and the \nvss\ radio galaxy survey are examined for correlations induced by the \isw\ effect.  
\wmap\ contains many receivers with different channels and frequency bands in order to eliminate foreground contamination and instrumental systematics from observations.  In order to enhance the signal-to-noise ratio, a co-added \wmap\ map is constructed from a noise weighted sum of the maps observed by different channels and receivers.   Residual Galactic emission and bright point sources are removed by the application of a mask.  It is this co-added \wmap\ map that is examined for correlation with the \nvss\ data. 
The large sky coverage and source distribution  of the \nvss\ data make it an ideal probe of the local matter distribution to use when searching for the \isw\ effect.  Sources in the catalogue are thought to be distributed in the redshift range $0<z<2$, with a peak distribution at $z\sim0.8$\cite{boughn:2002}.  This corresponds closely to redshift regions where the \isw\ signal is expected to be produced\cite{afshordi:2004}.  
Not all of the sky is sufficiently observed in the \nvss\ catalogue.  A joint mask is constructed to exclude from subsequent analyses those regions of the sky not observed in the catalogue, in addition to the regions excluded by the \wmap\ mask.  The \wmap\ and \nvss\ data, with the joint mask applied, are illustrated in \fig{\ref{fig:isw_maps}}.

Monte Carlo simulations are performed in order to constrain the statistical significance of any detection of correlation between the \wmap\ and \nvss\ data.  1000 simulations of the \wmap\ co-added map are constructed, modelling carefully the beam and anisotropic noise properties of each of the \wmap\ channels and mimicking the co-added map construction procedure.  No correlation exists between these \wmap\ simulations and the \nvss\ data, hence the analyses subsequently performed on the data may be repeated on the simulations and compared in order to constrain the significance of any correlation detected in the data.

\newlength{\mapplotwidth}
\setlength{\mapplotwidth}{75mm}

\begin{figure}
\centering
\subfigure[\wmap]{\includegraphics[viewport= 0 40 800 440,clip=,width=\mapplotwidth]{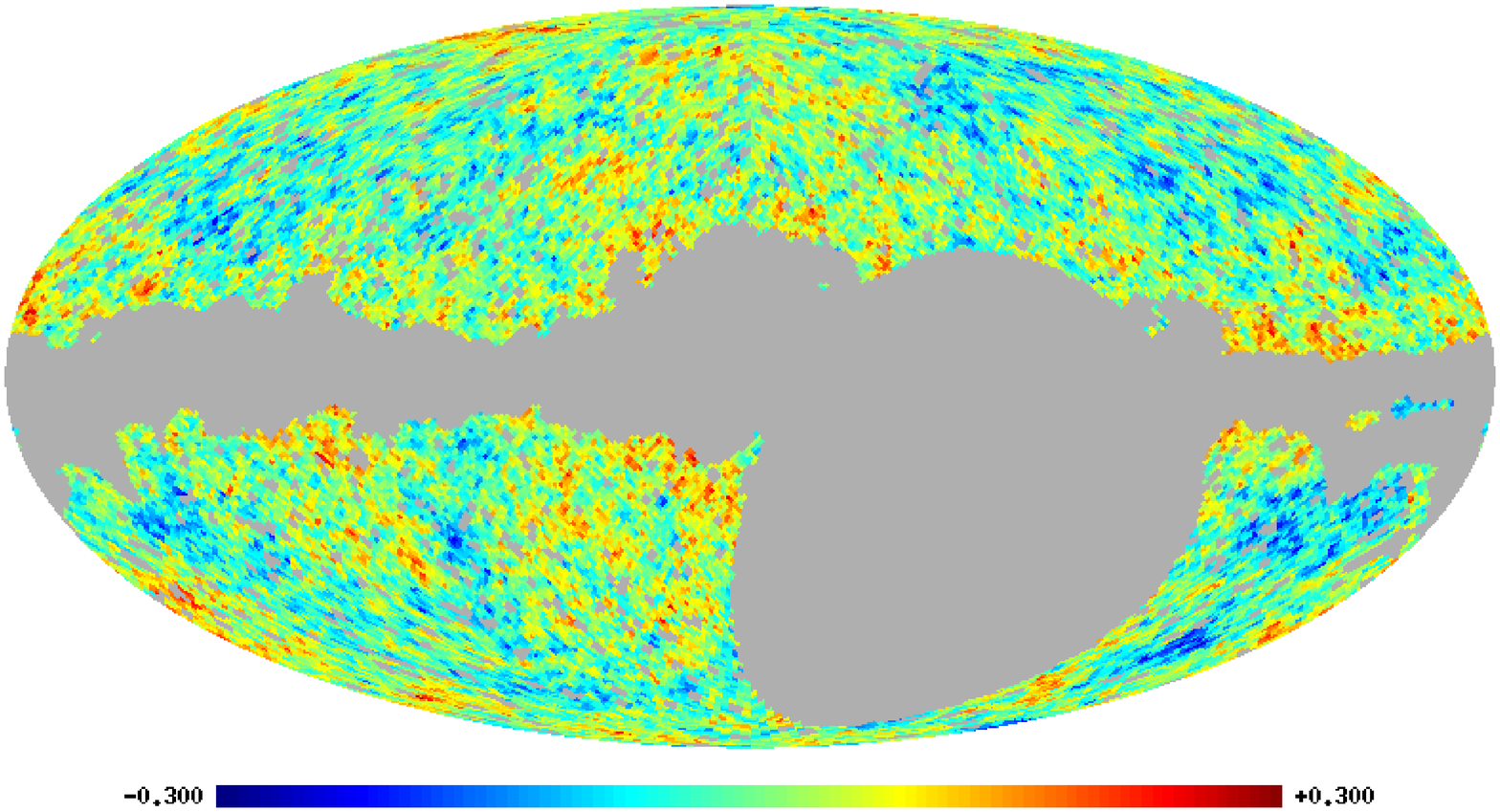}}
\subfigure[\nvss]{\includegraphics[viewport= 0 40 800 440,clip=,width=\mapplotwidth]{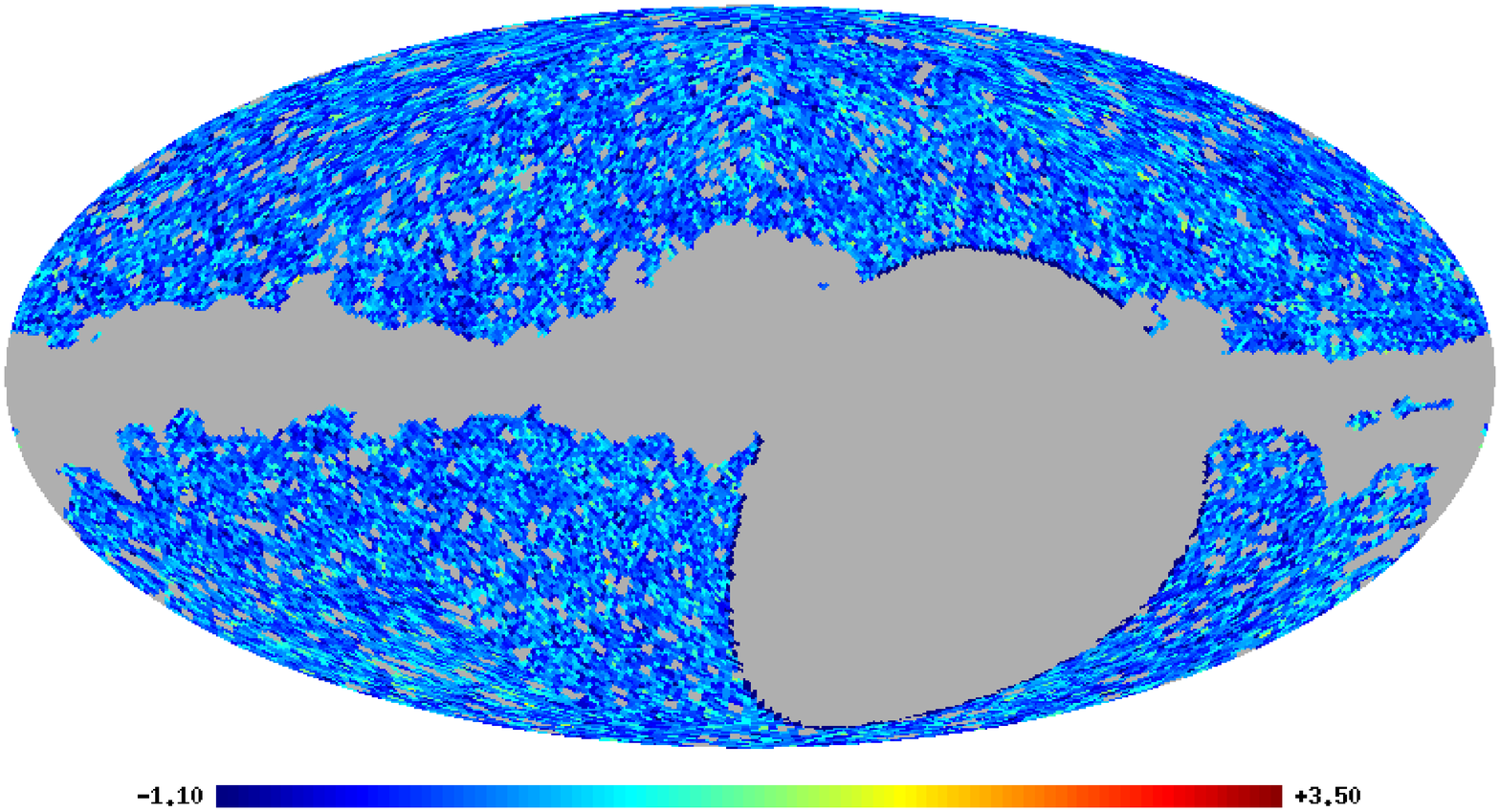}}
\caption{\wmap\ co-added three-year and \nvss\ maps after application of the joint mask.}
\label{fig:isw_maps}
\end{figure}

\subsection{Procedures} 

The analysis procedures consist of computing the wavelet correlation estimators described by \eqn{\ref{eqn:covest}} for the various cases.
In the analysis performed by Ref.~\citenum{vielva:2005} the correlation is computed from the wavelet coefficients of the co-added \wmap\ and \nvss\ data using the axisymmetric \smhw\ for a range of isotropic scales.  The correlation is computed from the wavelets coefficients of both the \smhw\ and the \sbw\ by Ref.~\citenum{mcewen:2006:isw}.  In this directional setting anisotropic dilations are adopted and a range of \eulc\ orientations are considered.  The steerable \stwogdw\ is applied in Ref.~\citenum{mcewen:2007:isw2} to compute morphological measures of local features, which are then correlated.
Any deviation from zero in any of these estimators is an indication of a correlation between the \wmap\ and \nvss\ data and hence a possible detection of the \isw\ effect.  An identical analysis is performed using the simulated co-added maps in place of the \wmap\ data in order to construct significance measures for any detections made.  Finally, for the analyses that do not examine the morphology of local features, it is possible to use detections of the \isw\ effect to constrain properties of dark energy.

\subsection{Results} 

Detections of the \isw\ effect and, consequently, evidence for the existence of dark energy are presented in this section.  We first concentrate on the wavelet coefficient correlations performed by Refs.~\citenum{vielva:2005,mcewen:2006:isw}, before turning our attention to the correlation of measures of the morphology of local features, facilitated by the steerable \stwogdw, performed by Ref.~\citenum{mcewen:2007:isw2}.

\subsubsection{Wavelet coefficient correlation} 

Highly statistically significant detections of the \isw\ effect have been made using the wavelet correlation approach first pioneered by Ref.~\citenum{vielva:2005}, and since extended to a directional analysis by Ref.~\citenum{mcewen:2006:isw}.  The distribution of the wavelet correlation statistics were found to be Gaussian distributed, hence the approximate significance of detections of correlation may be inferred directly from the number of standard deviations (denoted \nsigma) that the correlation observed in the data deviations from the mean value computed from Monte Carlo simulations.  \nsigma\ surfaces computed for the \smhw\ and the \sbw\ are shown in \fig{\ref{fig:nsigma}} in the anisotropic dilation space.  For both wavelets, the maximum correlation detected is made at a level of $\nsigma\simeq3.9$ on wavelet scales about $\scaleavect=(100,300)^{\rm T}$ arcminutes.\cite{mcewen:2006:isw}  It should be noted that the level quoted for these detections is based on an \emph{a posteriori} selection of the scale corresponding to the most significant detection.  Quantifying the significance in this manner is satisfactory, provided that the \emph{a posteriori} nature of the analysis is acknowledged.
The sign (positive) and the scale of the detected correlation are consistent with an \isw\ signal.  Moreover, foreground and instrumental systematics have been analysed in detail and were found not to be the responsible for the correlation detected with either wavelet.\cite{vielva:2005,mcewen:2006:isw}  As a final consistency check, the regions on the sky contributing most significantly to the correlation observed were localised and examined in closer detail.  These regions were determined not to be atypical and do not, in general, correspond to regions with particularly bright point sources that potentially could have contaminated the data.\cite{mcewen:2006:isw}  These tests strongly suggest that the correlations detected by Refs.~\citenum{vielva:2005,mcewen:2006:isw} are due to the \isw\ effect, thereby independently verifying the existence of dark energy.

\newlength{\nsigmaplotwidth}
\setlength{\nsigmaplotwidth}{70mm}

\begin{figure}
\centering
\subfigure[\smhw]
  {\includegraphics[clip=,width=\nsigmaplotwidth]{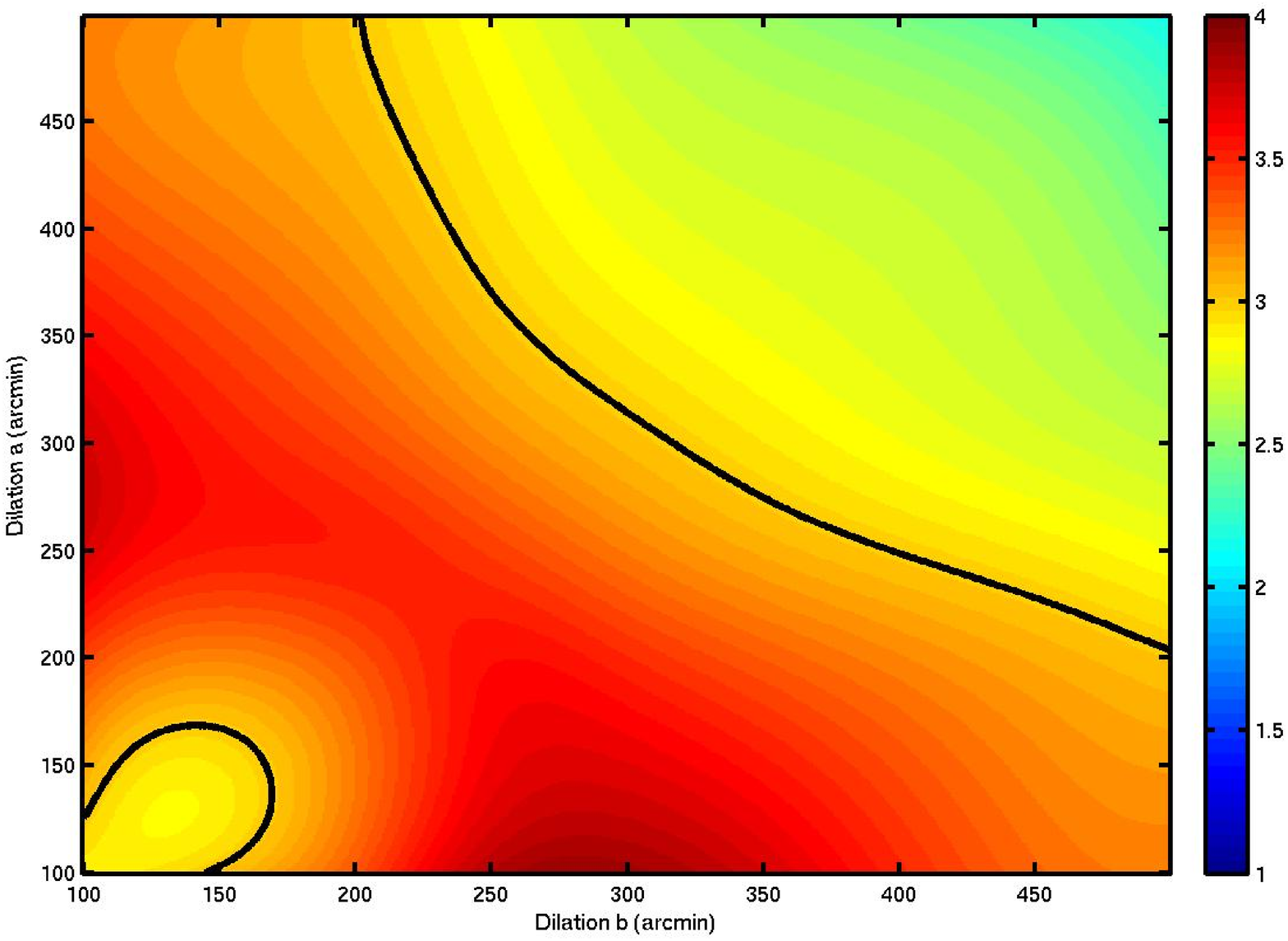}}
\subfigure[\sbw]
  {\includegraphics[clip=,width=\nsigmaplotwidth]{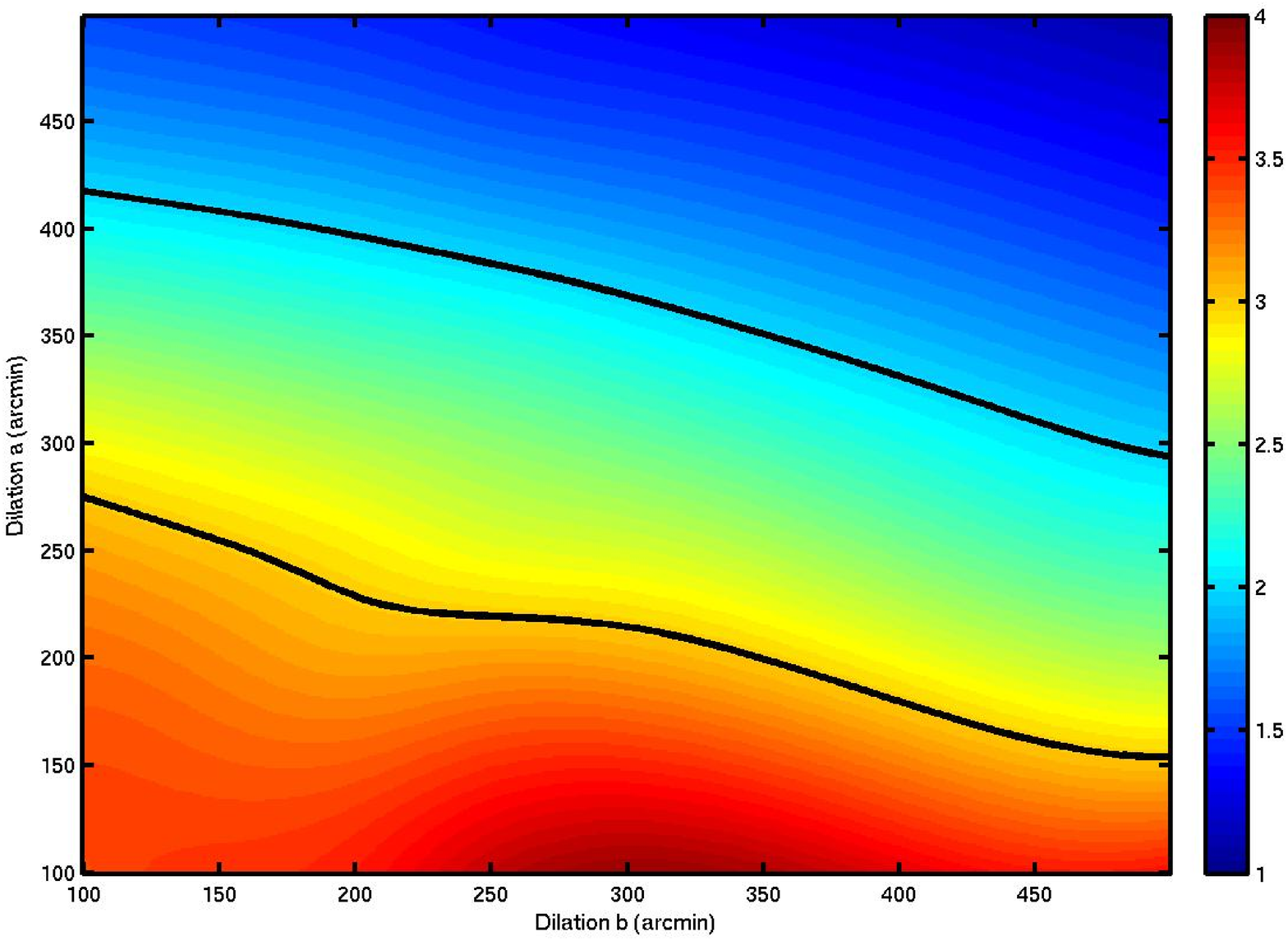}}
\caption{Wavelet correlation \nsigma\ surfaces.  Contours are shown for levels of two and three \nsigma.}
\label{fig:nsigma}
\end{figure}

With the \isw\ effect detected successfully, it is possible to use the detection to constrain cosmological parameter that describe dark energy, such as the proportional energy density \Denlambda\ and the equation of state parameter \w.  Current constraints on \Denlambda\ suggest that dark energy constitutes approximately 74\% of the energy density of the Universe, \ie\ $\Denlambda\simeq0.74$, while constrains on \w\ are consistent with a value of $-1$ (corresponding to the simple cosmological constant case).\cite{spergel:2006}
By comparing the wavelet correlation measured from the data with theoretical predictions\footnote{The theoretical wavelet correlation for isotropic dilations is derived by Ref.~\citenum{vielva:2005}, while the extension to anisotropic dilations is derived by Ref.~\citenum{mcewen:2006:isw}.} (which vary as a function of \Denlambda\ and \w), it is possible to compute the likelihood distribution for the parameters. 
In the case of uniform priors on the parameters, the likelihood is identical to the posterior distribution, 
and parameter estimates can be inferred from the likelihood directly.  In \fig{\ref{fig:likelihood}} the likelihood recovered using the \smhw\ is illustrated (the likelihood recovered using the \sbw\ differs slightly, although inferences drawn from it are consistent with the \smhw\ case).  Marginalised distributions for \Denlambda\ and \w\ are recovered from the joint likelihood, with parameter estimates of 
$\Denlambda=0.63_{-0.17}^{+0.18}$ 
and
$\w=-0.77_{-0.36}^{+0.35}$
computed from the mean of the marginalised distributions.  These parameter estimates are consistent with estimates made by numerous other analysis techniques and data sets.  Although wavelets perform very well when attempting to detect the \isw\ effect since one may probe different scales and positions, once all information is incorporated to compute the likelihood surface the performance of a wavelet analysis is comparable to other linear techniques, as expected.  The power of the wavelet analysis is the ability to make highly significant detections in the first instance.  Nevertheless, the constraints placed on dark energy parameters provide a coherent picture, further verifying the existence of dark energy.

\newlength{\pdfwidth}
\setlength{\pdfwidth}{21mm}

\begin{figure}
\centering
  \subfigure[Full likelihood surface]{\includegraphics[height=\pdfwidth]{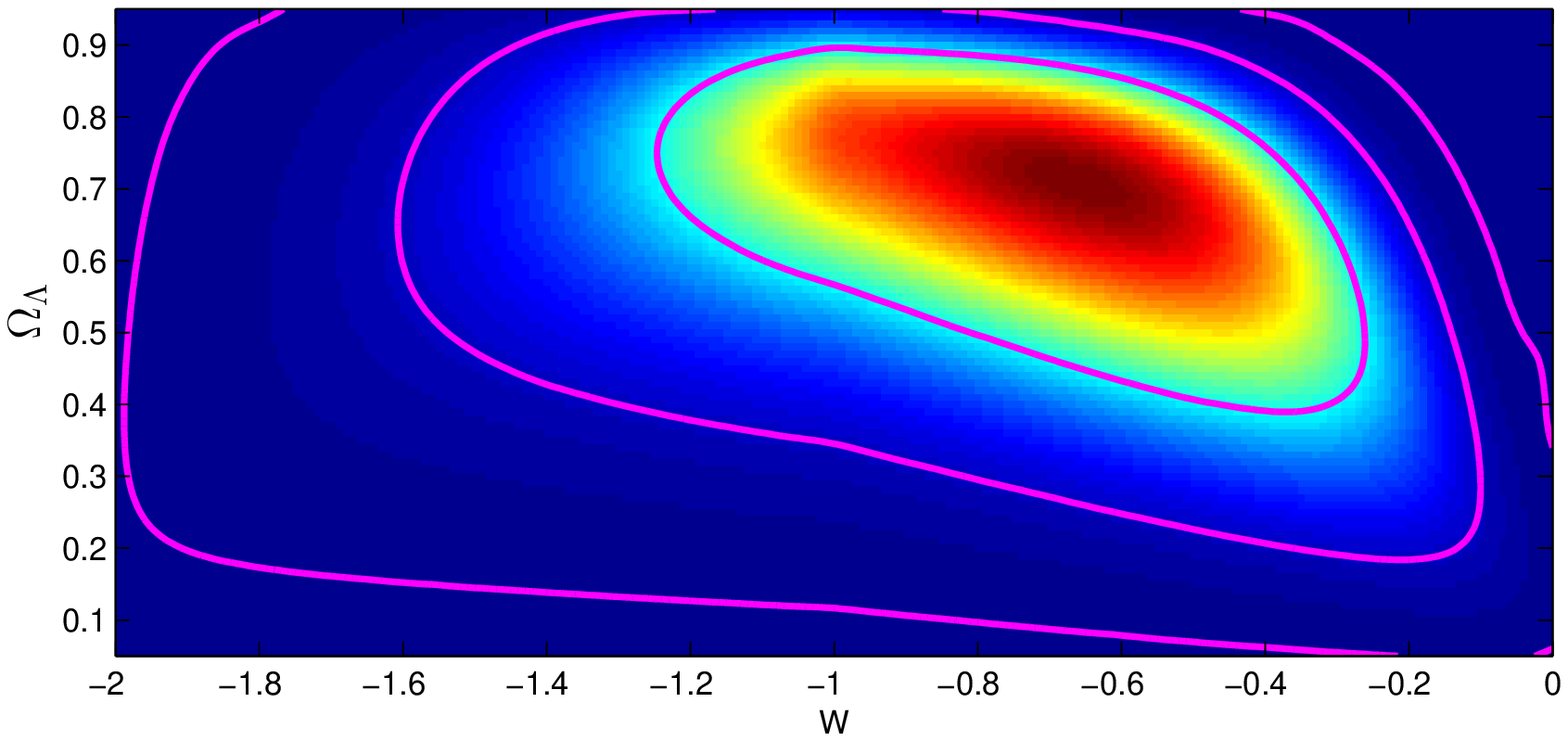}} \quad
  \subfigure[Marginalised distribution for \Denlambda]{\includegraphics[height=\pdfwidth]{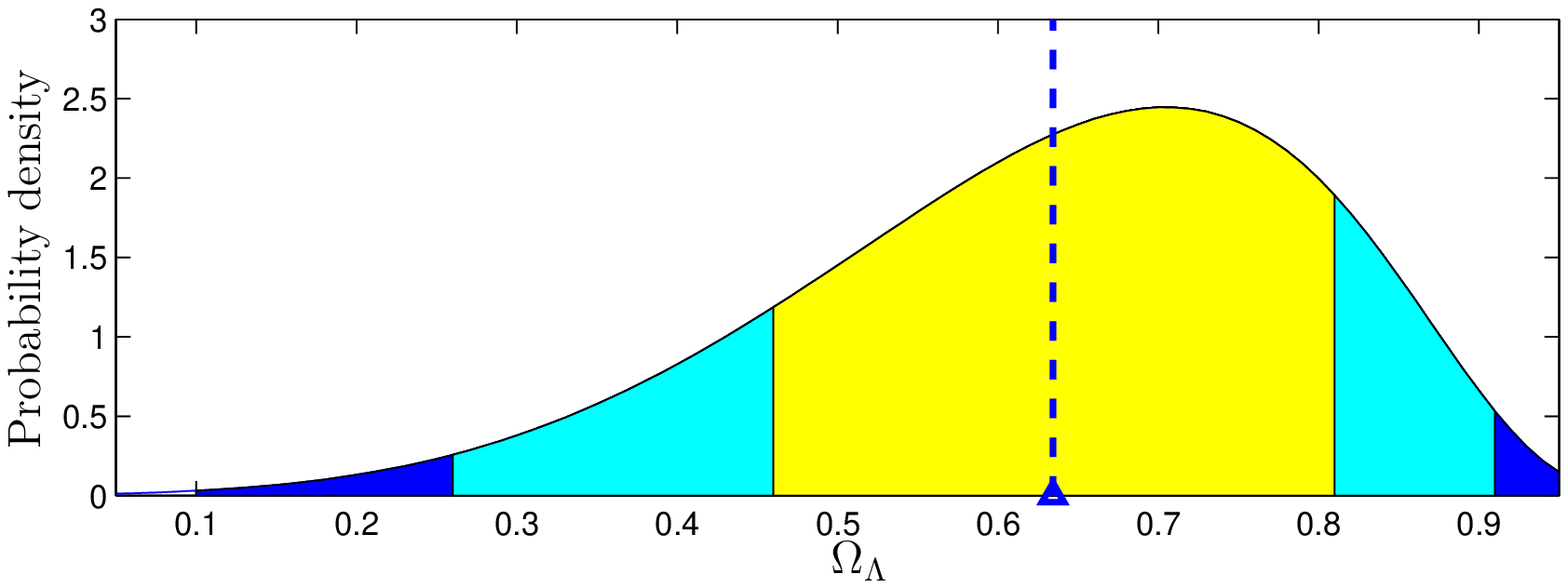}} \quad
  \subfigure[Marginalised distribution for \w]{\includegraphics[height=\pdfwidth]{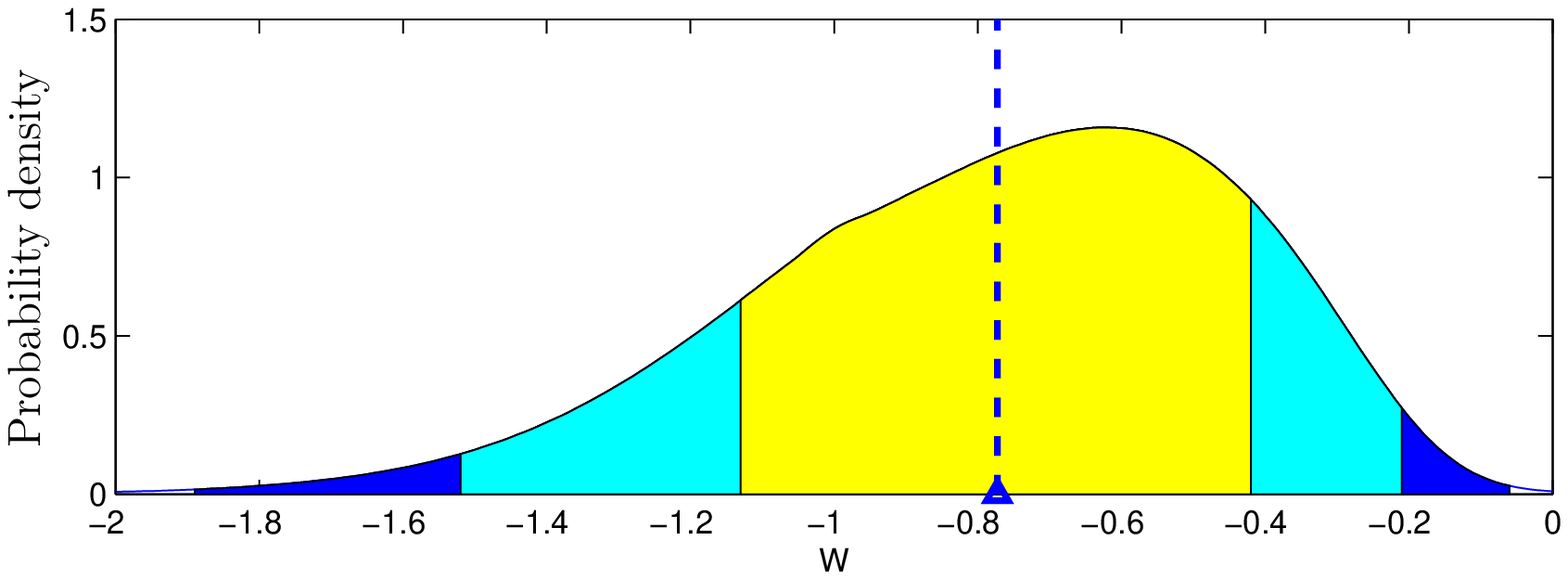}} \quad
\caption{Likelihoods constructed using the \smhw\ for parameters $(\Denlambda,\w)$.  The full likelihood surface is shown in panel~(a), with 68\%, 95\% and 99\% confidence contours also shown.  Marginalised distributions for each parameter are shown in the remaining panels, with 68\% (yellow/light-grey), 95\% (light-blue/grey) and 99\% (dark-blue/dark-grey) confidence regions also shown.  The parameter estimates made from the mean of the marginalised distribution are shown by the triangle and dashed line.}
\label{fig:likelihood}
\end{figure}

\subsubsection{Local morphological correlation}

A local morphological analysis has been performed by Ref.~\citenum{mcewen:2007:isw2} recently to search for the \isw\ effect, resulting in an alternative highly significant detection.  Rather than correlating the wavelet coefficients themselves, measures of the morphology of local features are correlated in this analysis, as described previously.  Not only does this probe the existence of the \isw\ effect, and hence dark energy, but it also provides insight into the morphological nature of the correlation induced between the \cmb\ and \lss.  This further insight may in future help to better understand the nature of dark energy.  The correlation of morphological measures of signed-intensity, orientation and elongation computed from the \wmap\ and \nvss\ data are displayed in \fig{\ref{fig:stat_morph}}, with significance levels obtained from Monte Carlo simulations also shown.  Only isotropic dilations are considered in this analysis.  A significant detection of correlation is made in the signed-intensity of local features, with the correlation computed from the data falling outside of the 99\% significance level on a scale of $\scalea=400$ arcminutes.  Moderate detections are also made in the other morphological measures (orientation and elongation), although the significance of these detections is lower.  In this analysis a \chisqd\ test was performed to examine the significance of the correlation when the statistics for all scales are considered in aggregate (as opposed to the \emph{a posteriori} analysis performed when considering a single scale only).  Using this test, correlation is detected in the signed-intensity of local features at 95\% significance.  On the scale ($\scalea=400$ arcminutes) corresponding to the most significant detection of correlation in the signed-intensity of local features, it may even be possible to observe the correlation by eye (see \fig{\ref{fig:wcoeff_maps}}).
Foregrounds and instrumental systematics were again examined and were determined not to be the source of the correlation signal observed.  The correlation detected indeed appears to be due to the \isw\ effect, thereby providing an alternative detection of dark energy.  In this setting, the use of the detection has not yet been used to constrain dark energy properties.
The analysis is complicated by allowing the orientation of wavelet coefficients to vary as a function of the data, hence the derivation of the theoretical correlation for each local morphological measure is not easily tractable.  The derivation of theoretical correlations and dark energy constrains in this setting is the focus of current research.

\newlength{\statplotwidth}
\setlength{\statplotwidth}{52mm}

\begin{figure}
\centering
\mbox{
\subfigure[Signed-intensity]{
\includegraphics[width=\statplotwidth]{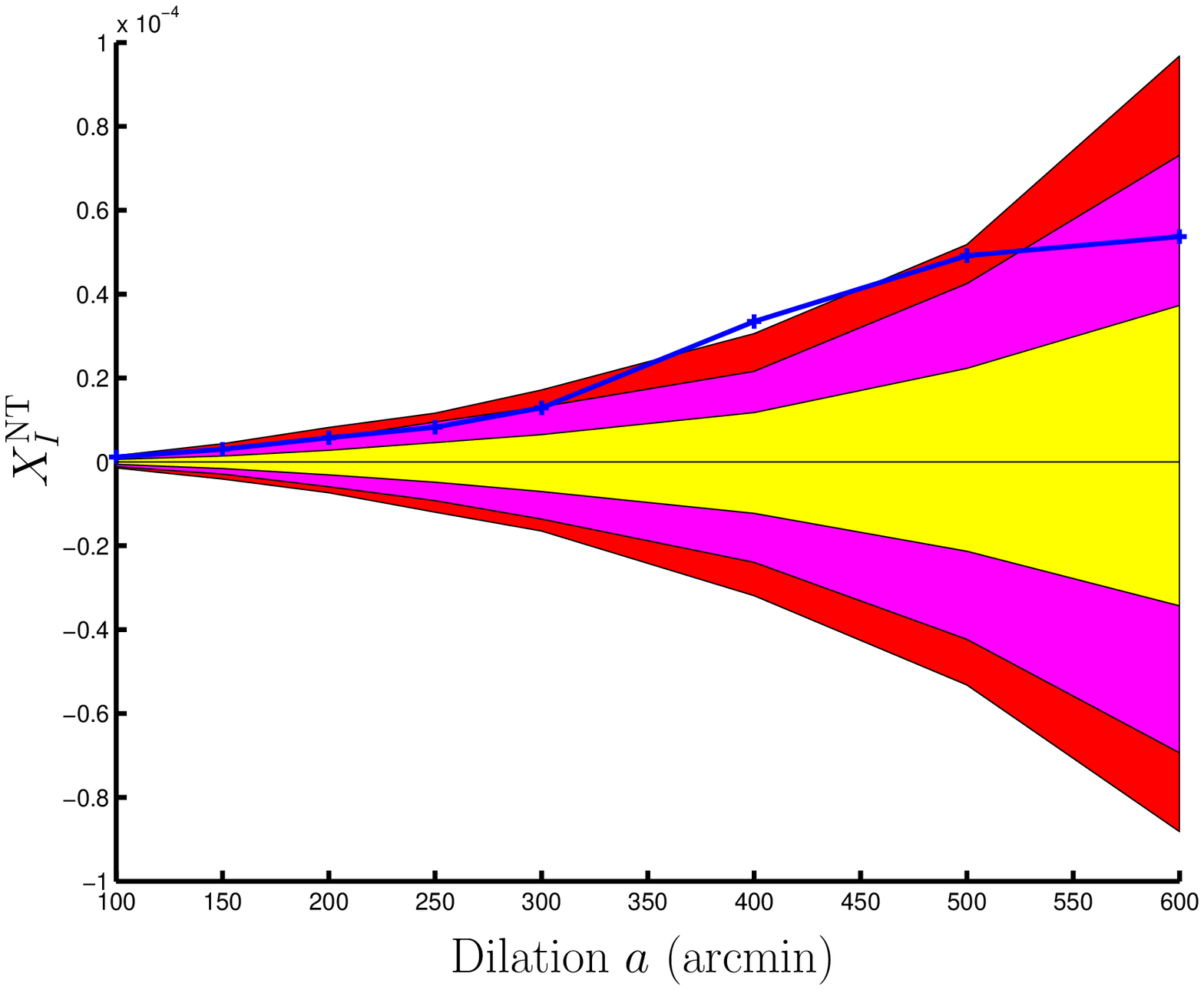}
}\quad
\subfigure[Orientation]{
\includegraphics[width=\statplotwidth]{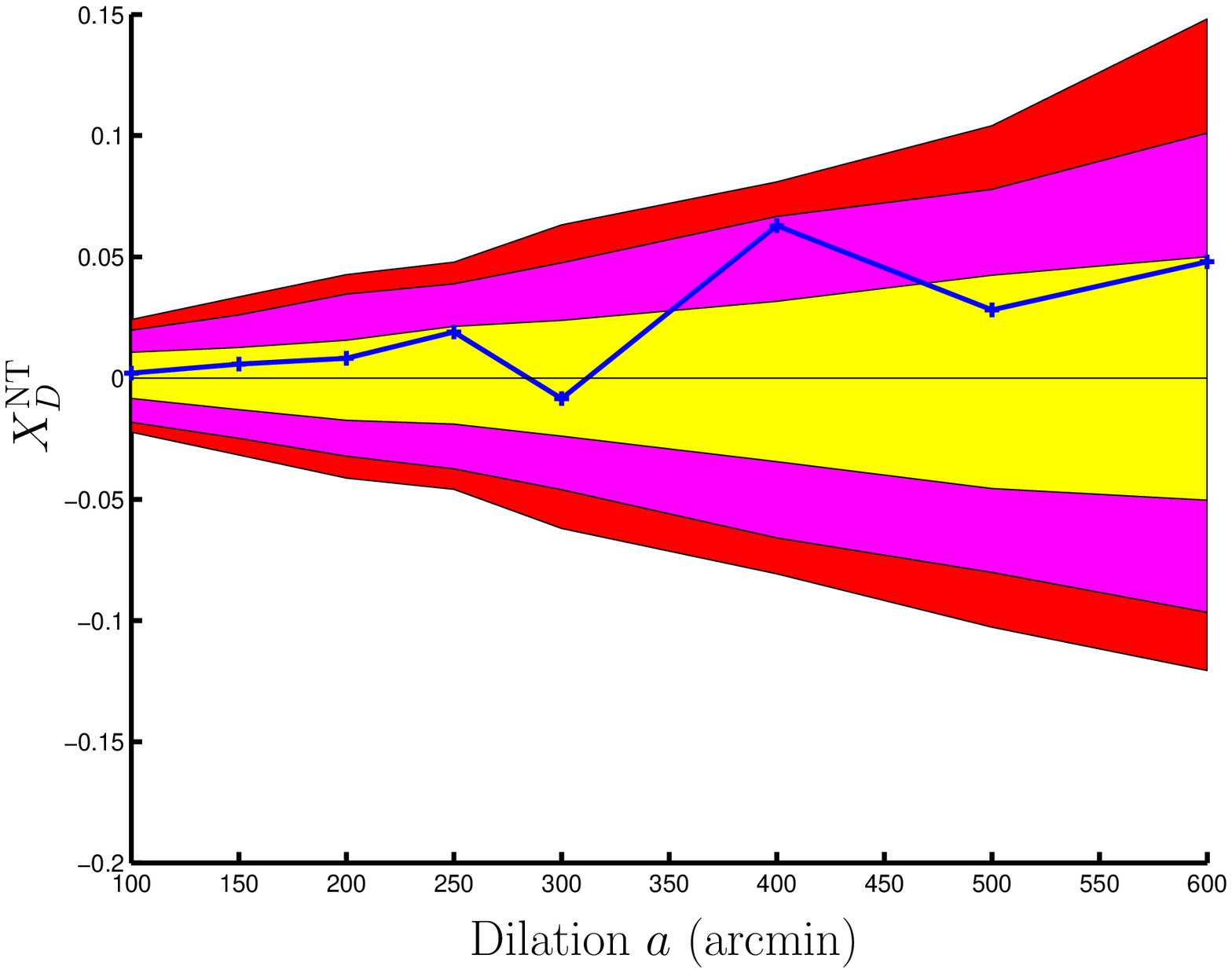}
}\quad
\subfigure[Elongation]{
\includegraphics[width=\statplotwidth]{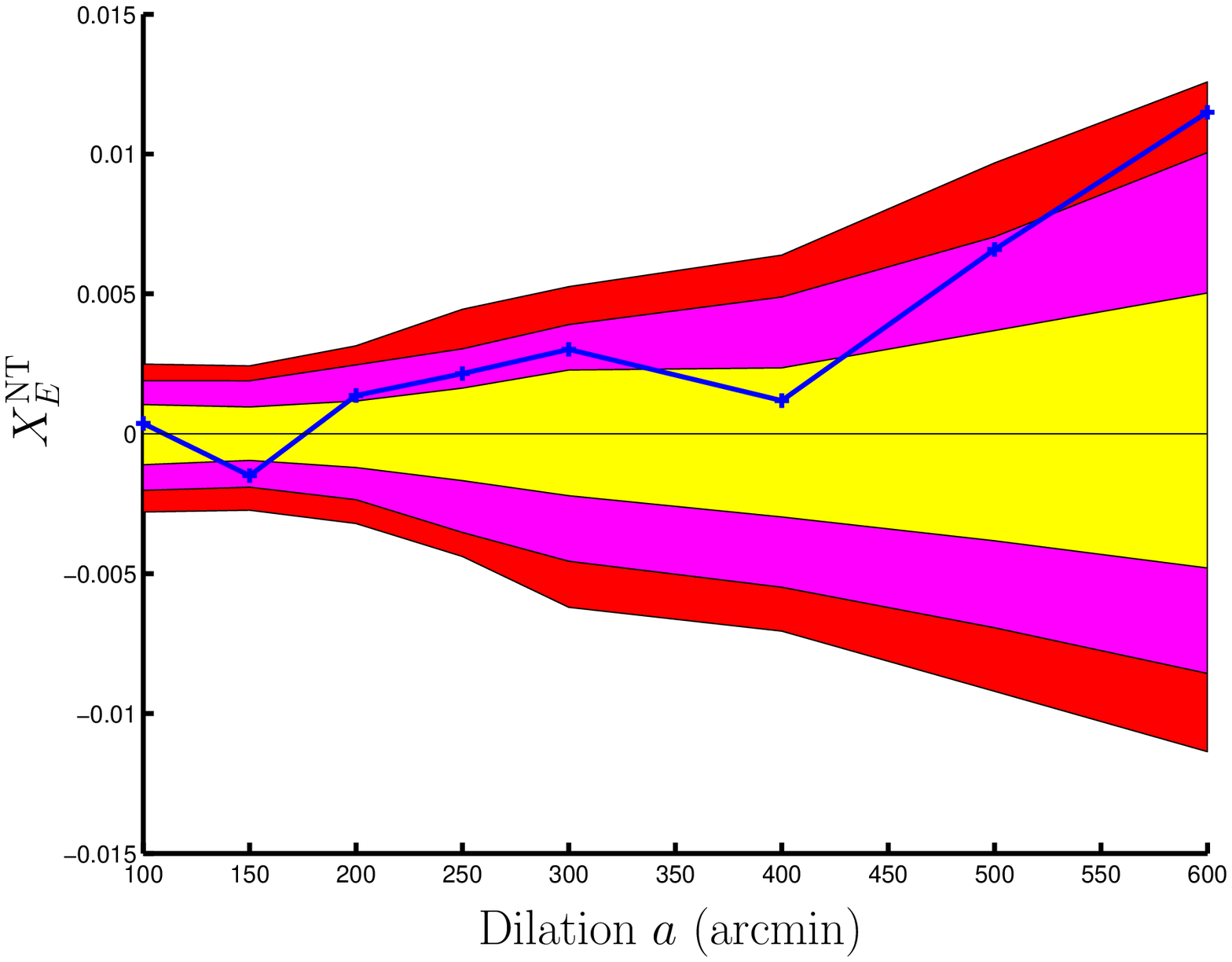}
}
}
\caption{Correlation statistics computed for each morphological measure. Significance levels obtained from 1000 Monte Carlo simulations are shown by the shaded regions for 68\% (yellow/light-grey), 95\% (magenta/grey) and 99\% (red/dark-grey) levels.}
\label{fig:stat_morph}
\end{figure}

\setlength{\mapplotwidth}{75mm}
\begin{figure}
\centering
\subfigure[\wmap]{\includegraphics[clip=,width=\mapplotwidth]{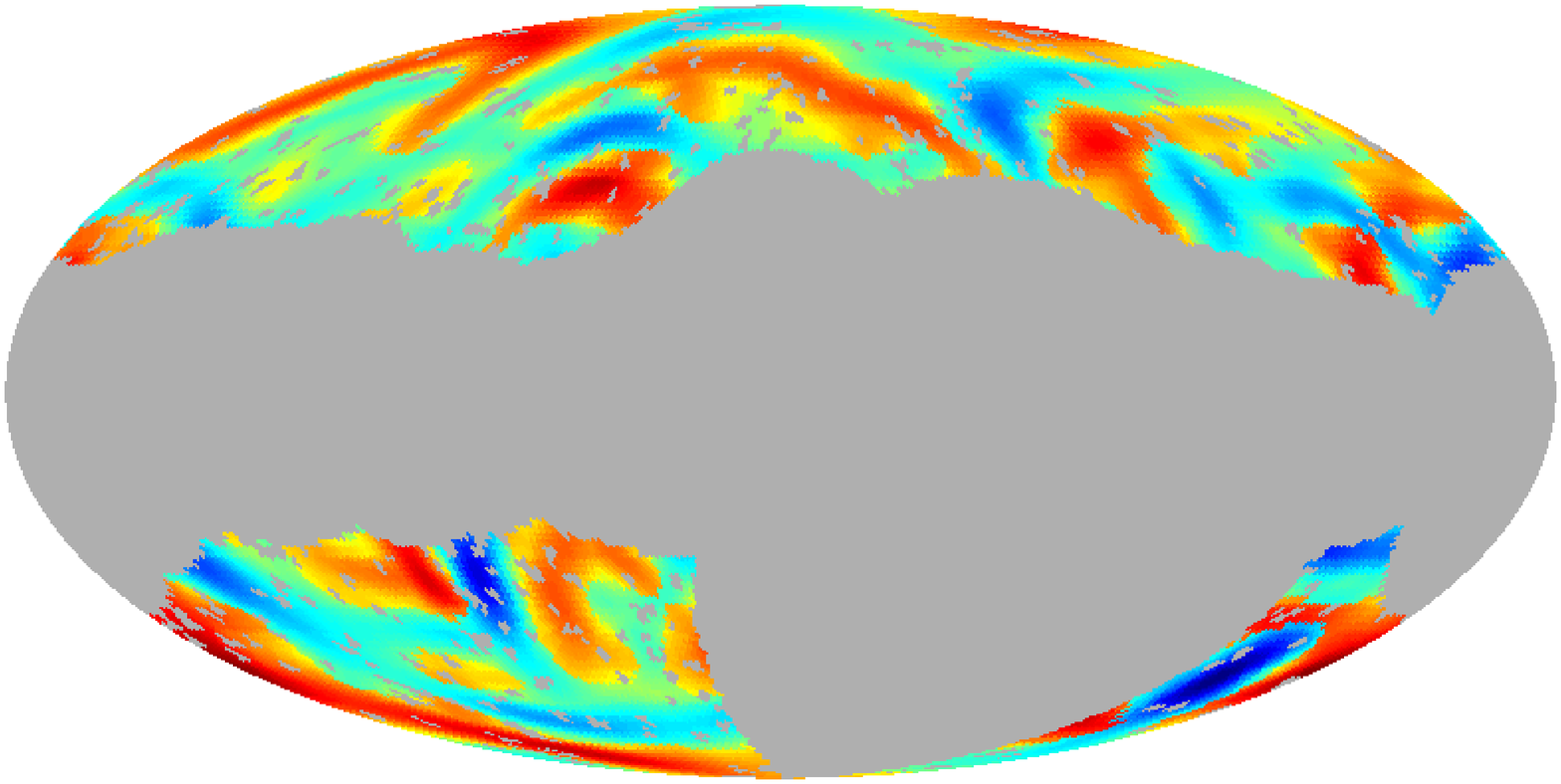}}
\subfigure[\nvss]{\includegraphics[clip=,width=\mapplotwidth]{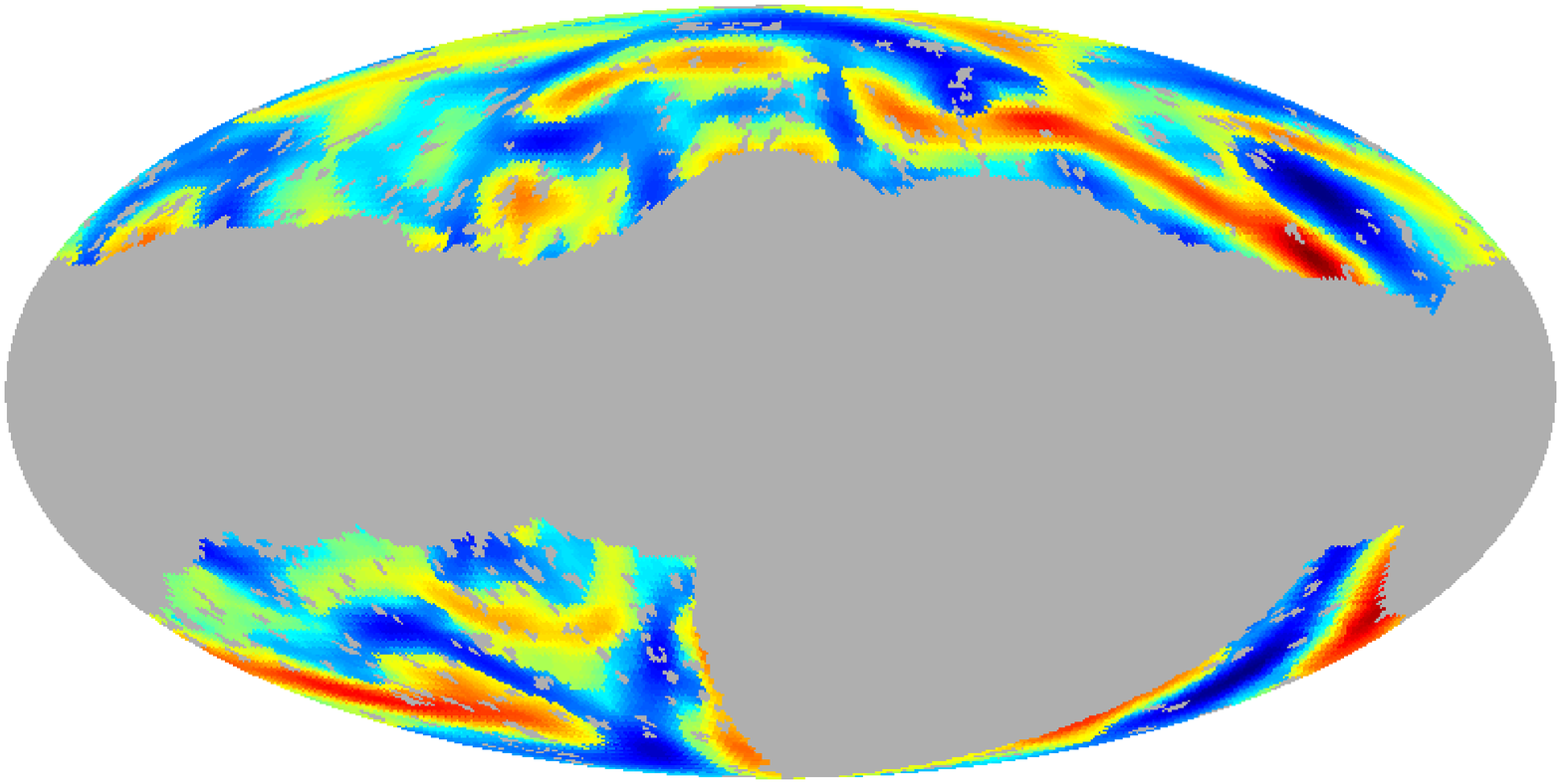}}
\caption{Morphological signed-intensity maps corresponding to the scale on which the maximum detection of correlation is made.  Arguably, it may be possible to observe some correlation between maps (a) and (b) by eye.
}
\label{fig:wcoeff_maps}
\end{figure}

\section{CONCLUSIONS} 
\label{sec:conclusions}

The \isw\ effect induces a correlation between the anisotropies of the \cmb\ and the \lss\ of the Universe that exists only in the presence of dark energy.  This provides an independent physical phenomenon that may be used to verify the existence of dark energy.  Moreover, the particular correlation signal induced by the \isw\ effect may, in certain cases, be used to constrain properties of dark energy.  The use of wavelets on the sphere to detect a correlation between the \wmap\ \cmb\ data and the \nvss\ radio galaxy survey has been presented.  Two approaches have been reviewed.  The first approach searches for a correlation between the wavelet coefficients of two data sets\cite{vielva:2005,mcewen:2006:isw}, whereas the second searches for a correlation between measures of the morphology of local features in the two data sets\cite{mcewen:2007:isw2}.  Positive detections of the \isw\ effect are made using both approaches, thereby independently verifying the existence of dark energy.  The effectiveness of a wavelet analysis on the sphere is demonstrated by the high statistical significant of the detections.  Various tests have shown that in all cases the correlation detected is indeed due to the \isw\ effect and not contamination in the data or instrumental effects.  Furthermore, for the first of the two techniques (the wavelet coefficient correlation), it is possible to use the correlation signal detected to constrain cosmological parameters that describe dark energy (using the correlation detected in the morphological analysis to constrain dark energy parameters is the focus of current research).  Parameter estimates have been obtained that are consistent with estimates made using numerous other analysis techniques and data sets.  The independent analyses performed therefore provide a coherent picture, further verifying the existence of dark energy.  

In more sophisticated dark energy models, perturbations in the dark energy fluid should be taken into account.  Wavelet based \isw\ detections could be used to probe this property and to constrain the corresponding sound speed of dark energy (see \eg\ Ref.~\citenum{weller:2003}).  This is an interesting area of future research, where again
the use of wavelets on the sphere to detect the \isw\ effect may help us to better understand the nature of dark energy, the dominant component of our Universe of which we know very little currently.

\acknowledgments     

This review is based on the works performed in collaboration with Patricio Vielva, Yves Wiaux, Mike Hobson, 
Enrique Mart{\'\i}nez-Gonz\'alez, Pierre Vandergheynst and Anthony Lasenby.  These works would not have been completed in the first instance without the contribution of all involved.  JDM would also like to extend additional thanks to Mike Hobson for comments on this manuscript.  This work has been supported by the Particle Physics and Astronomy Research Council, UK.


\bibliographystyle{spiebib}   
\bibliography{bib}   

\end{document}

%% file: thesis_macros.tex
\newcommand{\eqn}[1]{Eqn.~(#1)}

\newcommand{\fig}[1]{Fig.~#1}

\newcommand{\sectn}[1]{Sec.~#1}

\newcommand{\eg}{\mbox{\it e.g.}}
\newcommand{\ie}{\mbox{\it i.e.}}

\newcommand{\cmb}{{CMB}}
\newcommand{\cmbtext}{{cosmic microwave background}}

\newcommand{\wmap}{{WMAP}}
\newcommand{\wmaptext}{{Wilkinson Microwave Anisotropy Probe}}

\newcommand{\cobedmr}{\mbox{COBE-DMR}}
\newcommand{\cobedmrtext}{Cosmic Background Explorer-Differential Microwave Radiometer}
\newcommand{\cobe}{\mbox{COBE}}

\newcommand{\isw}{{ISW}}
\newcommand{\iswtext}{{integrated Sachs-Wolfe}}

\newcommand{\lss}{{LSS}}
\newcommand{\lsstext}{large scale structure}

\newcommand{\nvss}{{NVSS}}
\newcommand{\nvsstext}{{NRAO VLA Sky Survey}}

\newcommand{\lcdm}{\ensuremath{\Lambda}{CDM}}
\newcommand{\lcdmtext}{\ensuremath{\Lambda} Cold Dark Matter}

\newcommand{\smhw}{{SMHW}}

\newcommand{\sbw}{{SBW}}
\newcommand{\stwogdw}{{S2GDW}}

\newcommand{\spcend}{\ensuremath{\:}}

\newcommand{\cconj}{\ensuremath{\ast}} 
\newcommand{\reals}{\ensuremath{\mathbb{R}}}

\newcommand{\integers}{\ensuremath{\mathbb{Z}}}
\newcommand{\naturals}{\ensuremath{\mathbb{N}}}
\newcommand{\ltwo}{\ensuremath{\mathrm{L}^2}}
\newcommand{\sphere}{\ensuremath{{\mathbb{S}^2}}}
\newcommand{\sothree}{\ensuremath{{\mathrm{SO}(3)}}}
\newcommand{\vect}[1]{\ensuremath{\mbox{\boldmath ${#1}$}}}

\newcommand{\nsigma}{\ensuremath{N_\sigma}}

\newcommand{\opnexpv}{\ensuremath{\langle}}
\newcommand{\clsexpv}{\ensuremath{\rangle}}

\newcommand{\dx}{\ensuremath{\mathrm{\,d}}}
\newcommand{\dmu}[1]{\ensuremath{\dx \Omega(#1)}}

\newcommand{\deul}[1]{\ensuremath{\dx \varrho(#1)}}

\newcommand{\ctime}{\ensuremath{\eta}}

\newcommand{\den}{\ensuremath{\rho}}

\newcommand{\Den}{\ensuremath{\Omega}}

\newcommand{\Denlambda}{\ensuremath{\Den_{\Lambda}}}
\newcommand{\pres}{\ensuremath{p}}
\newcommand{\w}{\ensuremath{w}}
\newcommand{\z}{\ensuremath{z}}

\newcommand{\sa}{\ensuremath{\omega}}
\newcommand{\saa}{\ensuremath{\theta}}
\newcommand{\sab}{\ensuremath{\varphi}}
\newcommand{\sas}{\ensuremath{\saa, \sab}}
\newcommand{\eul}{\ensuremath{\mathbf{\rho}}}
\newcommand{\euls}{\ensuremath{\eula, \eulb, \eulc}}
\newcommand{\eula}{\ensuremath{\alpha}}
\newcommand{\eulb}{\ensuremath{\beta}}
\newcommand{\eulc}{\ensuremath{\gamma}}
\newcommand{\el}{\ensuremath{\ell}}
\newcommand{\m}{\ensuremath{n}}

\newcommand{\p}{\ensuremath{^\prime}}

\newcommand{\scalea}{\ensuremath{a}}
\newcommand{\scaleb}{\ensuremath{b}}
\newcommand{\scaleab}{\ensuremath{{\scalea,\scaleb}}}
\newcommand{\scaleavect}{\ensuremath{\vect{a}}}

\newcommand{\kron}[2]{\ensuremath{\delta_{{#1}{#2}}}}

\newcommand{\shf}[2]{\ensuremath{Y_{#1#2}}}

\newcommand{\shc}[3]{\ensuremath{{#1}_{{#2}{#3}}}}

\newcommand{\sbessel}[1]{\ensuremath{j_{#1}}}

\newcommand{\dil}{\ensuremath{\mathcal{D}}}
\newcommand{\dilsmall}{\ensuremath{d}}
\newcommand{\spo}{\ensuremath{\Pi}}
\newcommand{\rot}{\ensuremath{\mathcal{R}}}

\newcommand{\sky}{\ensuremath{f}}

\newcommand{\skywavi}{\ensuremath{{{{W}}_\wavi^\sky}}}

\newcommand{\wavi}{\ensuremath{\psi}}

\newcommand{\admissC}{\ensuremath{C}}

\newcommand{\admissCli}{\ensuremath{{\admissC}_\wavi^\el}}

\newcommand{\Lopi}{\ensuremath{{L}_\wavi}}

\newcommand{\cocycle}{\ensuremath{\lambda}}

\newcommand{\clnttheo}{\ensuremath{C_\el^{\rm NT}}}

\newcommand{\ndlab}{\ensuremath{{\rm N}}}
\newcommand{\tplab}{\ensuremath{{\rm T}}}
\newcommand{\dndz}{\ensuremath{\frac{\dx N}{\dx z}}}
\newcommand{\dgdz}{\ensuremath{\frac{\dx g}{\dx z}}}

\newcommand{\chisqd}{\ensuremath{\chi^2}}


%% file: isw_spie2007.bbl
\begin{thebibliography}{10}

\bibitem{hubble:1929}
E.~{Hubble}, ``A relation between distance and radial velocity among
  extra-galactic nebulae,'' {\em Proc.\ National Academy of Science}~{\bf 15},
  pp.~168--173, 1929.

\bibitem{sachs:1967}
R.~K. Sachs and A.~M. Wolfe, ``Perturbations of a cosmological model and
  angular variations of the microwave background,'' {\em Astrophys.\ J.}~{\bf
  147}, pp.~73--90, 1967.

\bibitem{spergel:2006}
D.~N. Spergel, R.~Bean, O.~Dor\'{e}, M.~R. Nolta, C.~L. Bennett, G.~Hinshaw,
  N.~Jarosik, E.~Komatsu, L.~Page, H.~V. Peiris, L.~Verde, C.~Barnes,
  M.~Halpern, R.~S. Hill, A.~Kogut, M.~Limon, S.~S. Meyer, N.~Odegard, G.~S.
  Tucker, J.~L. Weiland, E.~Wollack, and E.~L. Wright, ``Wilkinson {M}icrowave
  {A}nisotropy {P}robe ({WMAP}) three-year results: implications for
  cosmology,'' {\em Astrophys.\ J.} , in press, 2006.

\bibitem{crittenden:1996}
R.~G. Crittenden and N.~Turok, ``Looking for {$\Lambda$} with the
  {R}ees-{S}ciama effect,'' {\em Phys.\ Rev.\ Lett.}~{\bf 76}, pp.~575--578,
  1996.

\bibitem{boughn:2002}
S.~P. Boughn and R.~G. Crittenden, ``Cross-correlation of the cosmic microwave
  background with radio sources: constraints on an accelerating universe,''
  {\em Phys.\ Rev.\ Lett.}~{\bf 88}, p.~021302, 2002.

\bibitem{bennett:2003a}
C.~L. Bennett, M.~Halpern, G.~Hinshaw, N.~Jarosik, A.~Kogut, M.~Limon, S.~S.
  Meyer, L.~Page, D.~N. Spergel, G.~S. Tucker, E.~Wollack, E.~L. Wright,
  C.~Barnes, M.~R. Greason, R.~S. Hill, E.~Komatsu, M.~R. Nolta, N.~Odegard,
  H.~V. Peiris, L.~Verde, and J.~L. Weiland, ``First-year {W}ilkinson
  {M}icrowave {A}nisotropy {P}robe ({WMAP}) observations: preliminary maps and
  basic results,'' {\em Astrophys.\ J.\ Supp.}~{\bf 148}, p.~1, 2003.

\bibitem{hinshaw:2006}
G.~Hinshaw, M.~R. Nolta, C.~L. Bennett, R.~Bean, O.~Dor\'{e}, M.~R. Greason,
  M.~Halpern, R.~S. Hill, N.~Jarosik, A.~Kogut, E.~Komatsu, M.~Limon,
  N.~Odegard, S.~S. Meyer, L.~Page, H.~V. Peiris, D.~N. Spergel, G.~S. Tucker,
  L.~Verde, J.~L. Weiland, E.~Wollack, and E.~L. Wright, ``Three-year
  {W}ilkinson {M}icrowave {A}nisotropy {P}robe ({WMAP}) observations:
  temperature analysis,'' {\em Astrophys.\ J.} , in press, 2006.

\bibitem{boughn:2004}
S.~Boughn and R.~Crittenden, ``A correlation of the cosmic microwave sky with
  large scale structure,'' {\em Nature}~{\bf 427}, pp.~45--47, 2004.

\bibitem{condon:1998}
J.~J. Condon, W.~D. Cotton, E.~W. Greisen, Q.~F. Yin, R.~A. Perley, G.~B.
  Taylor, and J.~J. Broderick, ``The {NRAO} {VLA} sky survey,'' {\em
  Astrophys.\ J.}~{\bf 115}, p.~1693, 1998.

\bibitem{antoine:1998}
J.-P. Antoine and P.~Vandergheynst, ``Wavelets on the n-sphere and related
  manifolds,'' {\em J.\ Math.\ Phys.}~{\bf 39}(8), pp.~3987--4008, 1998.

\bibitem{pietrobon:2006}
D.~Pietrobon, A.~Balbi, and D.~Marinucci, ``Integrated {S}achs-{W}olfe effect
  from the cross-correlation of {WMAP} 3-year and {NVSS}: new results and
  constraints on dark energy,'' {\em ArXiv} , 2006.

\bibitem{vielva:2005}
P.~{Vielva}, E.~{Mart{\'{\i}}nez-Gonz{\'a}lez}, and M.~{Tucci},
  ``{Cross-correlation of the cosmic microwave background and radio galaxies in
  real, harmonic and wavelet spaces: detection of the integrated Sachs-Wolfe
  effect and dark energy constraints},'' {\em Mon.\ Not.\ Roy.\ Astron.\
  Soc.}~{\bf 365}, pp.~891--901, 2006.

\bibitem{barreiro:1997}
R.~B. {Barreiro}, J.~L. {Sanz}, E.~{Mart\'{\i}nez-Gonz\'{a}lez}, L.~{Cayon},
  and J.~{Silk}, ``{Peaks in the Cosmic Microwave Backgound: Flat versus Open
  Models},'' {\em Astrophys.\ J.}~{\bf 478}, pp.~1--+, 1997.

\bibitem{mcewen:2006:isw}
J.~D. McEwen, P.~Vielva, M.~P. Hobson, E.~Mart\'{\i}nez-Gonz\'{a}lez, and A.~N.
  Lasenby, ``Detection of the {ISW} effect and corresponding dark energy
  constraints made with directional spherical wavelets,'' {\em Mon.\ Not.\
  Roy.\ Astron.\ Soc.}~{\bf 373}, pp.~1211--1226, 2007.

\bibitem{mcewen:2007:isw2}
J.~D. McEwen, Y.~Wiaux, M.~P. Hobson, P.~Vandergheynst, and A.~N. Lasenby,
  ``Probing dark energy with steerable wavelets through correlation of {WMAP}
  and {NVSS} local morphological measures,'' {\em ArXiv} , 2007.

\bibitem{rubin:1980}
V.~C. {Rubin}, N.~{Thonnard}, and W.~K. {Ford}, Jr., ``{Rotational properties
  of 21 SC galaxies with a large range of luminosities and radii, from NGC 4605
  /R = 4kpc/ to UGC 2885 /R = 122 kpc/},'' {\em Astrophys.\ J.}~{\bf 238},
  pp.~471--487, 1980.

\bibitem{riess:1998}
A.~G. {Riess}, A.~V. {Filippenko}, P.~{Challis}, A.~{Clocchiatti},
  A.~{Diercks}, P.~M. {Garnavich}, R.~L. {Gilliland}, C.~J. {Hogan}, S.~{Jha},
  R.~P. {Kirshner}, B.~{Leibundgut}, M.~M. {Phillips}, D.~{Reiss}, B.~P.
  {Schmidt}, R.~A. {Schommer}, R.~C. {Smith}, J.~{Spyromilio}, C.~{Stubbs},
  N.~B. {Suntzeff}, and J.~{Tonry}, ``Observational evidence from supernovae
  for an accelerating universe and a cosmological constant,'' {\em Astron.\
  J.}~{\bf 116}, pp.~1009--1038, 1998.

\bibitem{perlmutter:1999}
S.~Perlmutter, G.~Aldering, G.~Goldhaber, R.~A. Knop, P.~Nugent, P.~G. Castro,
  S.~Deustua, S.~Fabbro, A.~Goobar, D.~E. Groom, I.~M. Hook, A.~G. Kim, M.~Y.
  Kim, J.~C. Lee, N.~J. Nunes, R.~Pain, C.~R. Pennypacker, R.~Quimby,
  C.~Lidman, R.~S. Ellis, M.~Irwin, R.~G. McMahon, P.~Ruiz-Lapuente, N.~Walton,
  B.~Schaefer, B.~J. Boyle, A.~V. Filippenko, T.~Matheson, A.~S. Fruchter,
  N.~Panagia, H.~J.~M. Newberg, and W.~J. Couch, ``Measurements of {O}mega and
  {L}ambda from 42 high-redshift supernovae,'' {\em Astrophys.\ J.}~{\bf 517},
  pp.~565--586, 1999.

\bibitem{allen:2002}
S.~W. {Allen}, R.~W. {Schmidt}, and A.~C. {Fabian}, ``{Cosmological constraints
  from the X-ray gas mass fraction in relaxed lensing clusters observed with
  Chandra},'' {\em Mon.\ Not.\ Roy.\ Astron.\ Soc.}~{\bf 334}, pp.~L11--L15,
  2002.

\bibitem{penzias:1965}
A.~A. {Penzias} and R.~W. {Wilson}, ``A measurement of excess antenna
  temperature at 4080 {Mc/s},'' {\em Astrophys.\ J.}~{\bf 142}, pp.~419--421,
  1965.

\bibitem{nolta:2004}
M.~R. Nolta, E.~L. Wright, L.~Page, C.~L. Bennett, M.~Halpern, G.~Hinshaw,
  N.~Jarosik, A.~Kogut, M.~Limon, S.~S. Meyer, D.~N. Spergel, G.~S. Tucker, and
  E.~Wollack, ``First-year {W}ilkinson {M}icrowave {A}nisotropy {P}robe
  ({WMAP}) observations: dark energy induced correlation with radio sources,''
  {\em Astrophys.\ J.}~{\bf 608}, pp.~10--15, 2004.

\bibitem{wiaux:2005}
Y.~Wiaux, L.~Jacques, and P.~Vandergheynst, ``Correspondence principle between
  spherical and {E}uclidean wavelets,'' {\em Astrophys.\ J.}~{\bf 632},
  pp.~15--28, 2005.

\bibitem{mcewen:2006:fcswt}
J.~D. McEwen, M.~P. Hobson, D.~J. Mortlock, and A.~N. Lasenby, ``Fast
  directional continuous spherical wavelet transform algorithms,'' {\em IEEE
  Trans.\ Sig.\ Proc.}~{\bf 55}(2), pp.~520--529, 2007.

\bibitem{wandelt:2001}
B.~D. Wandelt and K.~M. G\'{o}rski, ``Fast convolution on the sphere,'' {\em
  Phys.\ Rev.\ D.}~{\bf 63}(12), p.~123002, 2001.

\bibitem{wiaux:2006:review}
Y.~Wiaux, J.~D. McEwen, and P.~Vielva, ``Complex data processing: fast wavelet
  analysis on the sphere,'' {\em J.\ Fourier Anal.\ and Appl.} , invited
  contribution, in press, 2007.

\bibitem{afshordi:2004}
N.~Afshordi, ``Integrated {S}achs-{W}olfe effect in cross-correlation: the
  observer's manual,'' {\em Phys.\ Rev.\ D.}~{\bf D70}, p.~083536, 2004.

\bibitem{weller:2003}
J.~Weller and A.~M. Lewis, ``Large scale cosmic microwave background
  anisotropies and dark energy,'' {\em Mon.\ Not.\ Roy.\ Astron.\ Soc.}~{\bf
  346}, pp.~987--993, 2003.

\end{thebibliography}
